\documentclass[10pt]{iopart}

\usepackage{amsmath,amsfonts}
\usepackage{amsgen}
\usepackage{amssymb}
\usepackage{algorithmic}
\usepackage{algorithm}
\usepackage{array}
\usepackage{textcomp}
\usepackage{stfloats}
\usepackage{url}
\usepackage{verbatim}
\usepackage{graphicx}
\usepackage{cite}
\usepackage{bm}
\usepackage{siunitx}
\usepackage{algorithm}
\usepackage{verbatim}
\usepackage{caption}
\usepackage{subcaption}
\usepackage{multirow}
\usepackage{multicol}
\usepackage{comment}
\usepackage{upgreek}
\usepackage{rotating}
\usepackage{color}
\usepackage{ragged2e}
\usepackage{colortbl}
\usepackage{hhline}
\usepackage{titlesec}

\begin{document}

\title[\SUST]{Experimental Demonstration of a Superconductor SFQ-Based ADC for High-Frequency Signal Acquisition}

\author{B. Z. Ucpinar$^{1,2}$, S. Razmkhah$^{1,2}$, M. A. Karamuftuoglu$^{1,2}$, A. Bozbey$^{2}$}

\address{$^1$Ming Hsieh Department of Electrical and Computer Engineering, University of Southern California, Los Angeles, CA 90089 USA\\
$^2$Department of Electrical and Electronics Engineering and Computer Engineering, TOBB University of Economics and Technology, Ankara, Turkey\\}

\ead{ucpinar@usc.edu}

\vspace{10pt}

\begin{indented}
\item[]September 2025
\end{indented}

\begin{abstract}
Superconducting quantum interference devices (SQUIDs) are among the most sensitive sensors, offering high precision through their well-defined flux–voltage characteristics. Building on this sensitivity, we designed, fabricated, and experimentally demonstrated a superconducting single flux quantum (SFQ)–based analog-to-digital converter (ADC) capable of detecting small variations in input current signals at high frequencies and converting them into SFQ pulse trains. To improve robustness and reduce errors, the design incorporates a majority circuit and two types of counters: asynchronous toggle flip–flop–based and synchronous cumulative-based, at the cryogenic stage. The counter collects the SFQ pulse train and converts it into a binary number, simplifying downstream digital readout. The circuits were implemented using the AIST CRAVITY (QuFab) HSTP process and successfully tested in our cryocooler system, validating both the design methodology and operation. This approach helps build a fully integrated system that combines digital SQUID functionality with cryogenic readout circuits on a single chip.
\end{abstract}

\vspace{2pc}


\maketitle

\ioptwocol
\section{Introduction}
Analog-to-digital converters (ADCs) and readout circuits that achieve high speed, high sensitivity, and ultra–low power operation are fundamental to cryogenic electronic systems. These circuits are key enablers for quantum computing, cryogenic sensing, and superconducting detector applications~\cite{5972910, razmkhahBook}. Because power dissipation at cryogenic temperatures is severely constrained while system performance often demands gigahertz-scale bandwidths with low noise~\cite{holmes2021cryogenic, razmkhah2024challenges}, superconducting electronics have become a compelling technological solution. Their negligible resistive losses and picosecond switching capabilities allow them to deliver exceptional speed and energy efficiency in low-temperature operation~\cite{4277786}.

Superconducting ADCs combine high bandwidth, sensitivity, and energy efficiency—features that conventional CMOS circuits cannot attain at cryogenic temperatures. Architectures based on superconducting technology have demonstrated sampling frequencies beyond GHz levels and can directly digitize fast-varying analog signals with minimal energy loss.

Superconducting ADC implementations have evolved along several directions. Voltage-to-frequency or pulse-frequency modulation (PFM) ADCs use SFQ pulse counting to encode analog levels; for example, 9-bit superconducting DACs employing SFQ PFM techniques demonstrated high conversion rates using Nb Josephson integration technology \cite{mizugaki20149}. Sigma–Delta ADC architectures in superconducting technology utilize oversampling and noise shaping to convert analog inputs into SFQ pulse streams, thereby achieving high resolution and SNR \cite{guelaz2009wide}. Early SQUID-based quantizers introduced bidirectional pulse outputs—one for increasing and one for decreasing input currents—fed into counters for digitization, offering high linearity and sensitivity \cite{phillips1987superconducting}. Moreover, superconducting digitizers, such as voltage-to-frequency (V/F) converters and time-to-digital converters (TDCs), have been integrated with low-temperature detectors, achieving a time resolution of 30 ps and a full-scale current of sub-microamperes ($\mu$A) \cite{mukhanov20117}. Our early work also utilized the quasi one-junction SQUIDs (QOS) for converting the input from photon detectors to SFQ pulses \cite{9153926,9373955}. These existing approaches confirm that superconductors are well-suited for cryogenic, low-power, high-speed ADC tasks; however, many architectures remain relatively modular—i.e., ADC plus readout or post-processing—rather than fully integrated with error mitigation.

The motivation for this design stems from the limitations of conventional ADCs and readout methods in superconducting systems. While superconductors offer unmatched sensitivity, existing approaches often necessitate complex room-temperature readout circuits, which increase system cost and latency while limiting bandwidth. By integrating the ADC directly on-chip using a digital SQUID structure, our approach converts analog signals into binary data locally, enabling compatibility with on-chip digital signal processing (DSP) circuits. This integrated system increases the sampling frequency of the digital SQUID while eliminating the need for bulky external readout hardware, ultimately achieving enhanced speed, sensitivity, and efficiency for cryogenic applications.

Early implementations of SQUID systems faced significant limitations due to the complexity and cost of traditional DC SQUID readout circuits, which inherently restricted their dynamic readout speed. To address this, one of the first on-chip feedback-based architectures, where all circuit components were integrated on the same chip, was introduced in~\cite{Fujimaki1988_25}. Following this milestone, research on fully integrated SQUID-on-chip systems gained momentum.  

In \cite {Radparvar1994_26}, an alternative on-chip design was proposed that significantly improved data processing speed and eliminated the need for room-temperature signal processing electronics, while also supporting multi-channel operation. Similar integration-oriented approaches can be found in~\cite{Radparvar1997_27,Uhlmann1999_28}.  

Since a digital SQUID operates fundamentally as an ADC, several studies leveraged the high-precision $\Phi_{0}$-resolution ADC concepts reported in~\cite{Rylov1991_29}. Building upon this, a 500~MHz sampling-frequency digital SQUID was demonstrated in~\cite{Yuh1995_30}, and the first fully digital RSFQ-based implementation was presented in~\cite{Semenov2003_31}. These works showcased GHz-level data processing rates and architectures that perform digital quantization analogous to ADC behavior.  

In~\cite{Gupta2001_32}, it was shown that when the measured magnetic flux signals exhibit significant variations, a multi-SQUID approach---involving SQUIDs with different sensitivities and dynamic ranges---is advantageous. In this synchronous dual-SQUID system, one SQUID provides higher resolution while the other offers a wider operational range; the combined data are merged off-chip for post-processing.  

The study in~\cite{Reich2005_33} presented an SFQ-based digital SQUID architecture that simplifies circuit design without improving magnetic flux resolution, focusing instead on structural simplicity. Similarly,~\cite{Myoren2011_34} introduced a circuit configuration incorporating up/down counting registers to enhance the dynamic range of the digital SQUID. In this design, two distinct outputs from the SQUID are processed through up- and down-counters, enabling digital encoding of flux variations.  

To further improve the flux resolution ($\Phi_{0}$ sensitivity) of digital SQUIDs, on-chip feedback circuits were integrated in~\cite{Tsuga2013_35, Myoren2019_36}. These designs achieved a resolution improvement down to approximately $0.8\,\Phi_{0}$, marking a significant step forward in digital flux detection. However, despite such advancements, the resolution of digital SQUIDs still lags behind that of analog DC SQUIDs. Therefore, hybrid systems combining both digital SQUIDs and DC SQUIDs have been proposed, as demonstrated in~\cite{Reich2007_37}, to leverage the speed advantages of digital operation while maintaining the sensitivity of analog detection.

In this work, we present a fully integrated SFQ-based analog-to-digital converter (ADC) architecture that employs a pair of digital SQUID modulators with differing sensitivities to directly sample analog current signals and convert them into single-flux quantum (SFQ) pulse trains. The frequency of these pulses encodes the instantaneous variation of the input, with dual-rail quantization capturing both rising and falling transitions in the analog waveform. To enhance robustness and reduce bit error rate (BER), a majority-voting scheme is implemented, where three parallel modulators per ADC path generate independent quantized outputs, which are then combined through majority logic. The resulting SFQ pulse stream is processed by digital logic operating at cryogenic temperatures, producing a binary output compatible with downstream superconducting systems.

To extend the system's dynamic range and allow adaptive trade-offs between speed and precision, the two modulators, each with high and low sensitivity, respectively, are interfaced with a configurable digital backend. This backend consists of two signal processing units: an asynchronous counter optimized for high-speed, low-latency conversion, and a cumulative synchronous counter that integrates an up/down counting stage and averaging unit to improve resolution and noise resilience. A reconfigurable selection mechanism dynamically assigns each modulator output to either processing path, enabling flexible adaptation to varying input conditions or application requirements. The architecture offers runtime configurability and high-performance cryogenic analog-to-digital conversion across diverse operating regimes.

\begin{figure}[t]
\centering
\includegraphics[width=0.9\linewidth]{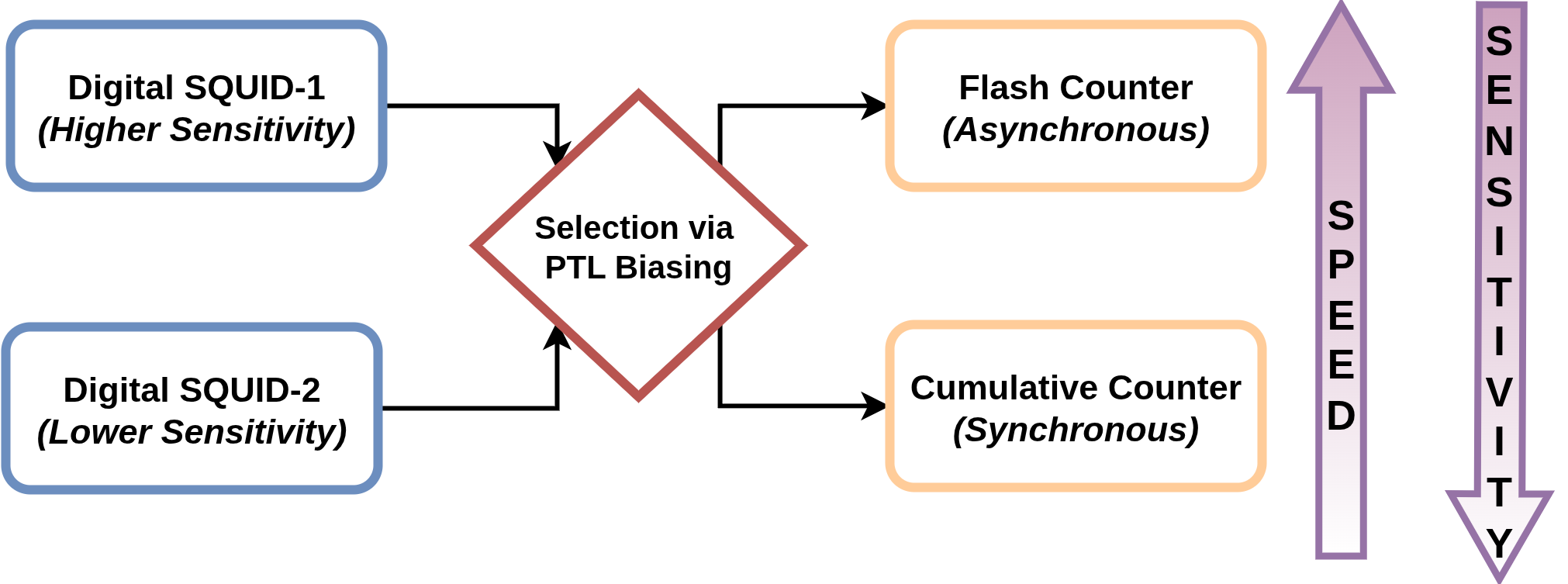}
\caption{Block diagram of the proposed SFQ-based ADC architecture.
The system comprises two digital SQUID modulators with distinct sensitivities and two digital processing units. The different sensitivities are achieved by tuning circuit parameters within a unified hardware platform. The processing units perform complementary functions: the flash counter is an asynchronous structure implemented with toggle flip-flops (TFFs) for rapid SFQ pulse counting, while the cumulative counter is a synchronous unit that includes an up/down counter and an averaging stage to enhance noise resilience. Selection between the modulators and processing units is managed through passive transmission line (PTL) links. Since PTLs do not convey bias currents, they define separate bias domains, allowing selective biasing of the desired circuit blocks.}
\label{fig:blockdiagram_sys}
\end{figure}

The remainder of this paper is organized as follows. Section~Theoretical Approach develops the quantization model and the statistical framework for error analysis, including the expected bit error rate (BER) reduction achieved through majority voting and a study of SFQ pulse dynamics. Section~Circuit Design presents schematic and simulation results for the digital SQUID modulator, majority logic, and SFQ counter circuits, along with fabrication details based on the AIST process~\cite{hidaka2021fabrication}. Section~Experimental Results describes the testbench configuration, measured waveforms, and verification of the design objectives. Section~Applications discusses integration within cryogenic systems, potential on-chip readout extensions, and application domains such as quantum sensing and high-speed data acquisition. Finally, Section~Conclusion summarizes the main findings and outlines directions for future work.

\section{Theoretical Approach}
\subsection{Superconducting Quantum Interference Devices (SQUIDs)}

A Superconducting Quantum Interference Device (SQUID) operates as a transducer that converts magnetic flux into voltage. It is regarded as one of the most sensitive magnetometers ever developed, capable of detecting changes in magnetic flux as small as $10^{-6}\Phi_0$. Because any physical quantity that can be transformed into magnetic flux—such as electrical current—can be measured by a SQUID, these devices find broad use in magnetometry \cite{5740577}, amplification \cite{Razmkhah_2021}, and the readout of highly sensitive quantum devices such as qubits \cite{Shnyrkov_2023}.

Analog SQUIDs are generally divided into two main categories: radio-frequency (RF) and direct-current (DC) types. The RF SQUID, introduced by Silver and Zimmerman \cite{zimmerman1966coherent}, consists of a single Josephson junction (JJ) integrated within a superconducting loop. The loop is inductively coupled to a resonant tank circuit composed of an inductor and a capacitor. An RF current drives the circuit, and magnetic flux variations are inferred from changes in the tank circuit’s resonance frequency. The DC SQUID, by contrast, contains two identical JJs connected in parallel within a superconducting loop. Its sensitivity, flux range, and operating characteristics are primarily determined by the critical current ($I_c$) of the junctions and the loop inductance \cite{clarke2006squid}.

Since our modulator design is based on the DC SQUID topology, the RF case will not be discussed further. In a DC SQUID, the bias current entering the superconducting loop divides between the two Josephson junctions. The current through each junction depends on its critical current and on the phase difference induced by the screening current that responds to the external magnetic flux threading the loop. The total current in the device is given by
\begin{equation}
I = I_1 + I_2 = I_c \sin(\theta_1) + I_c \sin(\theta_2),
\label{eq:dc_squid_current_initial}
\end{equation}
where $\theta_1$ and $\theta_2$ denote the superconducting phase differences across the junctions.  
The phase difference between the two junctions is related to the magnetic flux $\Phi$ through the loop by
\begin{equation}
\theta_2 - \theta_1 = 2n\pi + \frac{2\pi\Phi}{\Phi_0},
\label{eq:dc_squid_phase}
\end{equation}
where $\Phi_0 = h/(2e) = 2.07 \times 10^{-15}\,\text{Wb}$ is the magnetic flux quantum.  
Substituting this relation into (\ref{eq:dc_squid_current_initial}) yields the total current as
\begin{equation}
I = 2I_c \cos\left( \frac{n\Phi}{\Phi_0} \right) 
    \sin\left(\theta_1 + \frac{\pi\Phi}{\Phi_0}\right).
\label{eq:dc_squid_current}
\end{equation}

The total magnetic flux $\Phi$ in the superconducting loop is the sum of two contributions: the externally applied flux $\Phi_{\text{ext}}$ and the flux generated by the circulating screening current, $\Phi_{\text{loop}}$:
\begin{equation}
\Phi = \Phi_{\text{ext}} + \Phi_{\text{loop}}.
\label{eq:total_flux}
\end{equation}
The maximum current that can flow through the SQUID while both junctions remain superconducting is then given by
\begin{equation}
I_{\text{max}} = 2I_c 
\left|\cos\left( \frac{\pi\Phi_{\text{ext}}}{\Phi_0} \right)\right|.
\label{eq:max_current}
\end{equation}
Equation~(\ref{eq:max_current}) shows that the critical current reaches a maximum when $\Phi_{\text{ext}}/\Phi_0$ is an integer ($n \in \mathbb{N}$) and approaches zero when it is a half-integer ($n = (2m+1)/2$).  

Despite their remarkable sensitivity, conventional DC SQUIDs suffer from complex readout requirements. They typically rely on low-noise amplifiers, lock-in detectors, and feedback electronics—collectively forming a flux-locked loop (FLL)—that operate at room temperature. These external components not only increase system cost and complexity but also introduce additional noise and latency when transferring signals between the cryogenic and room-temperature domains, limiting the achievable readout speed.

\subsection{Digital SQUID}

\begin{figure}[!htbp]
 \centering
 \includegraphics[width=1\linewidth]{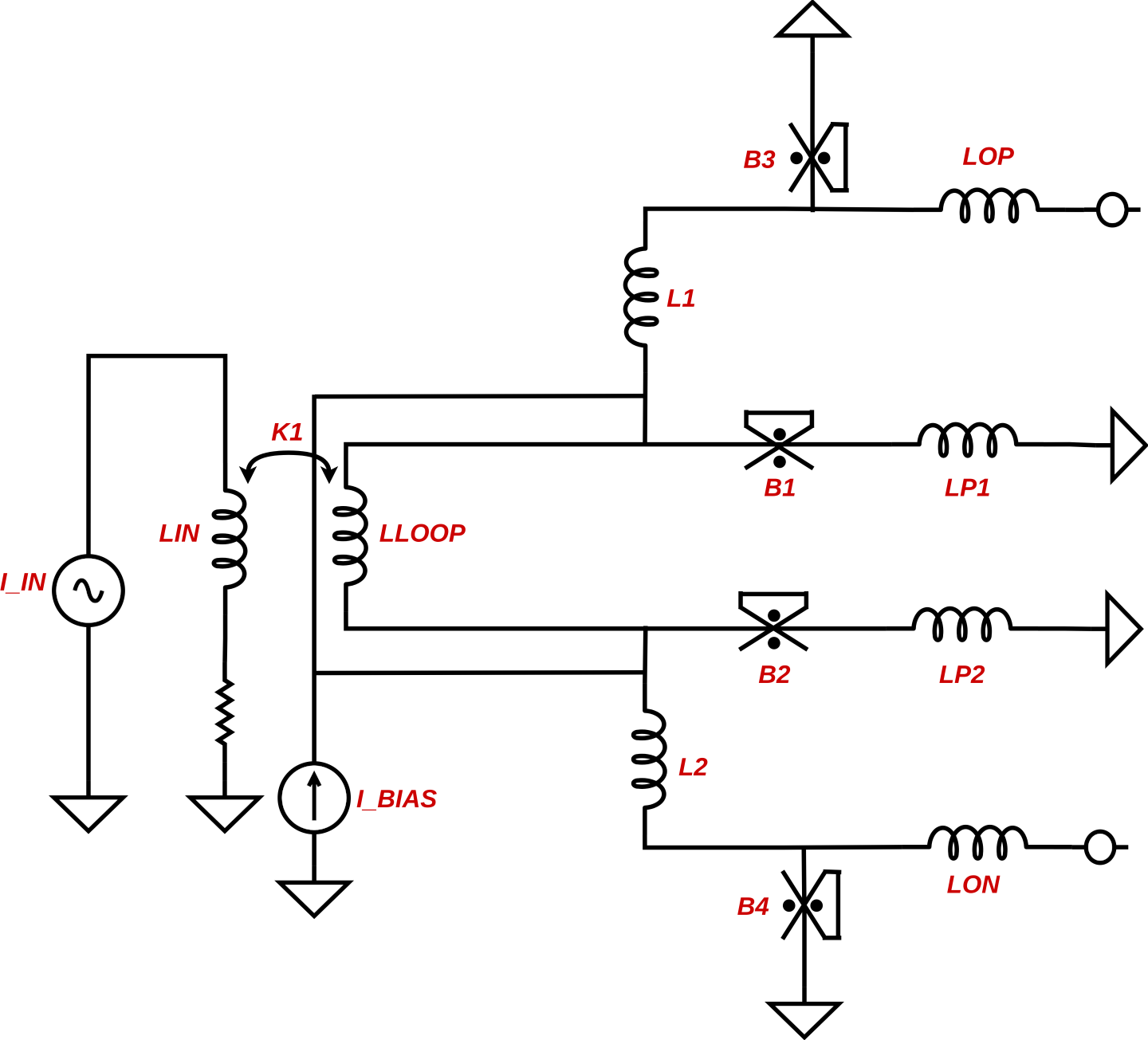}
 \caption{Circuit schematic of the digital SQUID designed using parameters from the HSTP process.
 The superconducting loop contains two Josephson junctions and a single inductor.
 Depending on the direction of the circulating current, either B1 or B2 switches release an SFQ pulse that relaxes the loop energy.
 The resulting current propagates through L1 or L2 and activates B3 or B4, which drive the digital processing stage.
 Each junction is resistively shunted so that $I_cR_n = 0.4$.}
 \label{fig:digiSquid_sch}
 \end{figure}

A digital SQUID operates on the same underlying Josephson-interference principle as a DC SQUID, but it reports magnetic flux in a quantized form rather than as a continuous voltage.
 Instead of a smooth voltage–flux characteristic, the device emits a sequence of single-flux-quantum (SFQ) pulses.
 Each pulse corresponds to the switching of one of the Josephson junctions (JJs) within the loop and reflects the direction of the applied magnetic field~\cite{5634076}.
 In effect, the circuit counts individual flux quanta that thread the loop, performing an analog-to-digital conversion of the magnetic signal coupled from a pick-up coil.
 The implementation used in this study is shown schematically in Fig.~\ref{fig:digiSquid_sch}.
Conventional DC SQUIDs can detect magnetic-flux variations as small as $10^{-6}\Phi_0$, but this extreme sensitivity requires elaborate, room-temperature readout electronics.
 Digital SQUIDs, in contrast, trade fine resolution for simplicity and speed.
 Because the output is inherently quantized in units of the flux quantum $\Phi_0$, the circuit operates without external feedback amplifiers or analog bias control.
 Its compact two-junction structure occupies little area, dissipates minimal power, and supports a broad dynamic range.
 When finer precision is needed, the digital SQUID can be paired with a DC SQUID in a hybrid configuration to recover sub-quantum sensitivity~\cite{yilmaz2019flux}.
Each JJ switching event represents a 2$\pi$ phase change and produces an SFQ pulse whenever the total flux in the loop changes by one $\Phi_0$.
 The device runs asynchronously—no global clock is required—and provides two complementary outputs that indicate positive or negative flux transitions.
The performance of a digital SQUID is mainly governed by the junction critical current ($I_c$) and the loop inductance ($L_{\text{loop}}$).
 The circulating-current increment associated with a single flux quantum is
 \begin{equation}
 \delta I = \frac{\Phi_0}{L_{\text{loop}}},
 \label{eq:current_change}
 \end{equation}
 showing that a larger inductance improves flux resolution.
 However, increasing $L_{\text{loop}}$ or lowering $I_c$ reduces the stored magnetic energy, making the device more sensitive to noise.
 The small-signal voltage response can be approximated by
 \begin{equation}
 \frac{\delta\langle V\rangle}{\delta\Phi_{\text{ext}}}
 = \frac{R_D,\delta I}{0.5\Phi_0}
 = \frac{2R_D}{L_{\text{loop}}},
 \label{eq:voltage_sensitivity}
 \end{equation}
 where $R_D$ is the dynamic resistance of the junction.
To ensure stable operation, the magnetic energy in the loop must greatly exceed the thermal energy:
 \begin{equation}
 \frac{\Phi_0^2}{2L_{\text{loop}}} \gg \frac{1}{2}k_BT.
 \label{eq:thermal_noise}
 \end{equation}
 This condition restricts how small $I_c$ and how large $L_{\text{loop}}$ can be chosen.
 A common design rule is
 \begin{equation}
 I_cL_{\text{loop}} > \Phi_0,
 \label{eq:flux_storage}
 \end{equation}
 which guarantees that flux quanta remain stably trapped in the loop and that random thermally activated switching is suppressed.

\subsection{Error Calculation}
One of the most relevant performance metrics for digital SQUIDs is the bit-error rate (BER), which expresses the probability of missing or falsely detecting an SFQ pulse.  
Errors originate primarily from Johnson noise and thermal fluctuations.  
Formally, the BER is written as
\begin{equation}
\text{BER} = \frac{N_{\text{errors}}}{N_{\text{total}}},
\end{equation}
where $N_{\text{errors}}$ is the number of incorrect or missed pulses and $N_{\text{total}}$ is the total number transmitted.

The chance of an erroneous switching event can be estimated from the ratio of thermal energy to the Josephson energy barrier using Kramers’ escape-rate model:
\begin{equation}
P_e \approx \exp\!\left(-\frac{E_J}{k_B T}\right),
\end{equation}
where $E_J$ is the Josephson energy, $k_B$ is Boltzmann’s constant, and $T$ is temperature.  
At high clock rates, additional error arises from timing jitter in SFQ pulse generation.  
Consequently, the total BER depends on both thermal activation and temporal jitter, each of which must be minimized to maintain reliable digital operation under cryogenic conditions.

\section{Circuit Design and Simulation}

The digital SQUID produces SFQ pulses proportional to the rate of change of magnetic flux.  
To suppress random errors, several SQUIDs can be operated in parallel, and their outputs combined through a majority-voting network, implementing triple-modular redundancy (TMR).  
Transmitting raw SFQ pulses directly to room-temperature electronics would require extremely wide bandwidths and could lead to pulse loss on long interconnects.  
To avoid this, on-chip RSFQ counters are used to accumulate pulses locally before they are sent off-chip.  
All components are integrated on a single die operating at $4.2\,\text{K}$, providing fast, low-latency readout with minimal power dissipation.

\subsection{Modulator Design}

The modulator circuit is based on a digital SQUID structure; however, unlike a conventional SQUID that detects magnetic flux through an external antenna, it is inductively coupled to an analog input signal.  
In this configuration, the device does not receive a direct electrical input but is driven through mutual coupling between two inductors—the input coil ($L_C$) and the loop inductance ($L_{IN}$).  
This coupling allows the magnetic flux generated by the input current to influence the superconducting loop.  
The Josephson junctions (JJs) inside the loop switch depending on the time derivative of this input current.  
When the current increases (positive slope), one of the JJs switches, producing a single-flux-quantum (SFQ) pulse at the positive output port.  
Conversely, when the current decreases (negative slope), the other JJ switches and emits a pulse at the negative port.  
Each switching event represents the emission of a single SFQ pulse accompanied by a $2\pi$ phase slip across the junction, corresponding to a quantized change in magnetic energy within the loop.

The overall sensitivity of the modulator is defined by the magnetic coupling coefficient ($K$) between $L_C$ and $L_{IN}$.  
A higher value of $K$ increases the magnetic flux linkage per unit of input current, which enhances the device’s ability to detect small variations in the analog signal.  
However, stronger coupling also narrows the dynamic range by causing earlier switching under large inputs.  
To investigate this balance between sensitivity and operating range, two modulator designs were implemented and simulated using different loop inductance values.

\subsubsection*{Modulator-1}

The design parameters for the first modulator (Modulator-1) are provided in Table~\ref{tab:squid1_params}.  
In this design, the superconducting loop inductance ($L_{loop}$) is 20\,pH, and the critical current ($I_c$) for both JJs (J1 and J2) is 216\,$\mu$A.  
This parameter set satisfies the flux-storage condition $2I_c L_{loop} > \Phi_0$, since  
$2 \times 216~\mu\text{A} \times 20~\text{pH} = 8.6 \times 10^{-15}~\text{V}\cdot\text{s} > \Phi_0$.  
The JJ parameters were selected to ensure that the modulator remains compatible with the biasing and signal levels of our RSFQ logic library.  
A 1.6\,mV bias voltage is applied to the circuit, supplying current to J1 and J2 through a pair of 50\,$\Omega$ resistors.  
The analog input current flowing through $L_{IN}$ is magnetically coupled to the loop inductance ($L_{LOOP}$), such that changes in the input flux cause redistribution of the circulating current.  
Depending on the direction of this induced current, either J1 or J2 switches to release an SFQ pulse accompanied by a $2\pi$ phase shift.  
These pulses are transferred to the next stage of digital logic through the output inductors L1 and L2, which serve as coupling elements to the RSFQ readout circuitry.

\begin{table}[!htbp]
\centering
\caption{Modulator-1 Design Values}
\label{tab:squid1_params}
\begin{tabular}{|c|c|c|c|}
\hline
\textbf{Parameter} & \textbf{Value} & \textbf{Parameter} & \textbf{Value} \\ \hline
$L_{IN}$ & 10\,pH & $L_{LOOP}$ & 20\,pH \\ \hline
$K_1$ & 0.5 & $L_{1}$ & 2.031\,pH \\ \hline
$L_{2}$ & 2.031\,pH & $L_{OP}$ & 2.031\,pH \\ \hline
$L_{ON}$ & 2.031\,pH & $J_1$, $J_2$, $J_3$, $J_4$ & 216\,$\mu$A \\ \hline
$L_{P1}$ & 0.086\,pH & $L_{P2}$ & 0.086\,pH \\ \hline
\end{tabular}
\end{table}

\subsubsection*{Modulator-2}

To enhance the system’s magnetic sensitivity, a second modulator variant (Modulator-2) was developed with a larger loop inductance, as summarized in Table~\ref{tab:squid2_params}.  
In this version, $L_{loop}$ was increased from 20\,pH to 100\,pH.  
A larger loop inductance decreases the current increment required to insert one flux quantum into the loop, thereby increasing the device’s ability to detect smaller flux changes and improving its overall sensitivity.  
However, this also means that the circuit stores less magnetic energy per flux quantum, which slightly increases its vulnerability to thermal noise and limits the highest achievable operating frequency.  
All other design parameters were kept the same to isolate the effect of the loop inductance on performance.

\begin{table}[!htbp]
\centering
\caption{Modulator-2 Design Values}
\label{tab:squid2_params}
\begin{tabular}{|c|c|c|c|}
\hline
\textbf{Parameter} & \textbf{Value} & \textbf{Parameter} & \textbf{Value} \\ \hline
$L_{IN}$ & 10\,pH & $L_{LOOP}$ & 100\,pH \\ \hline
$K_1$ & 0.5 & $L_{1}$ & 2.031\,pH \\ \hline
$L_{2}$ & 2.031\,pH & $L_{OP}$ & 2.031\,pH \\ \hline
$L_{ON}$ & 2.031\,pH & $J_1$, $J_2$, $J_3$, $J_4$ & 216\,$\mu$A \\ \hline
$L_{P1}$ & 0.086\,pH & $L_{P2}$ & 0.086\,pH \\ \hline
\end{tabular}
\end{table}

\subsection{Modulator Simulation Results}

\subsubsection{Modulator-1 Simulation Results}

\begin{figure}[!htbp]
\centering
\includegraphics[width=1\linewidth]{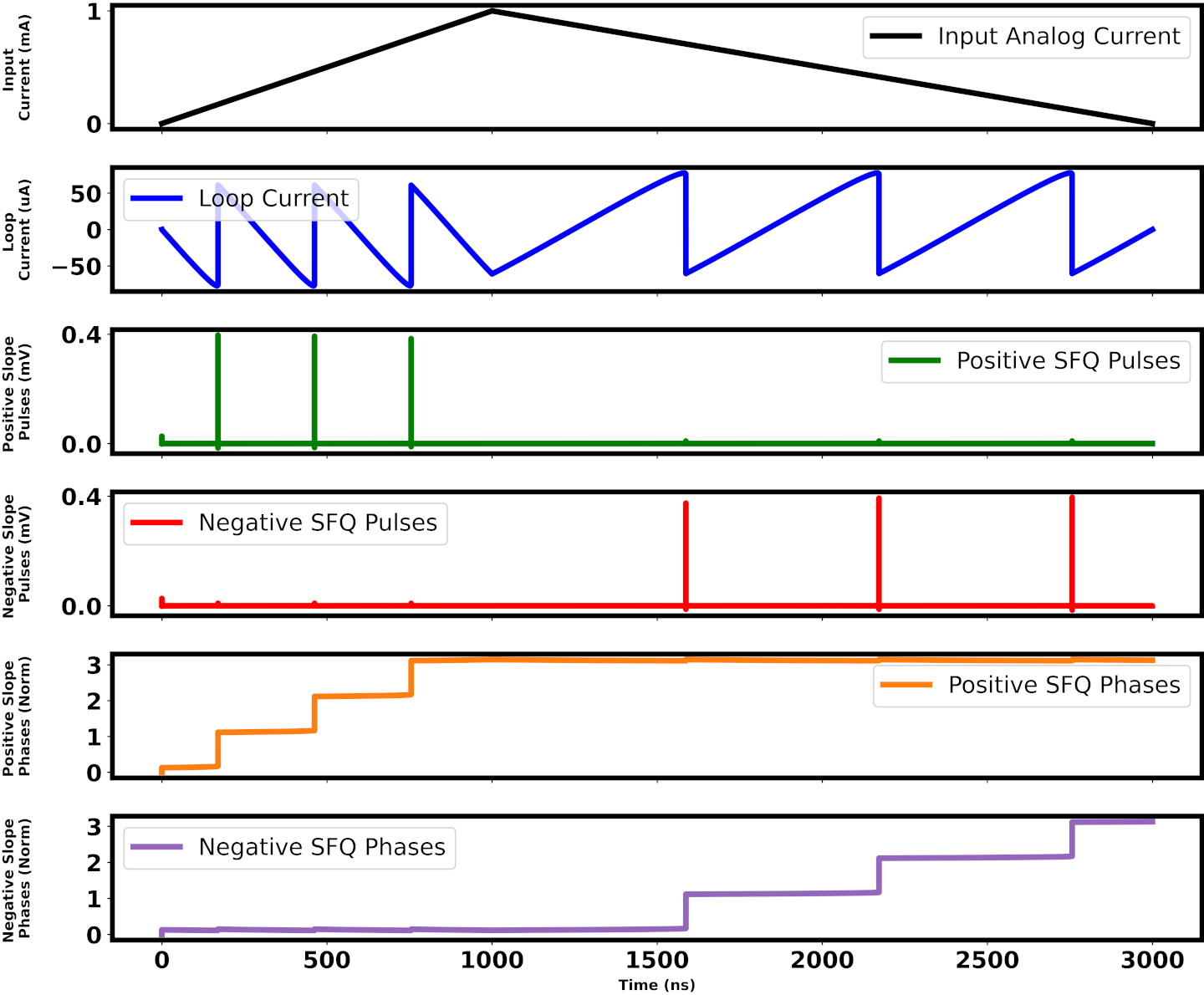}
\caption{Simulation result for Modulator-1.  
The input signal induces a circulating current in the loop, generating a screening current that opposes the applied flux.  
Once the screening current reaches its critical limit, one of the junctions switches to release an SFQ pulse, reducing the stored magnetic energy.  
Each switching corresponds to a quantized $2\pi$ phase change and represents one SFQ pulse transmitted to the subsequent logic stage.  
The pulse timing depends on the input slope: a steeper input slope produces faster pulse emission.  
In this example, let $m_1$ and $m_2$ denote the positive and negative input slopes, respectively, with $m_1 = 2m_2$.  
As a result, the time interval between two negative-slope pulses ($\Delta t_2$) is twice the interval between positive-slope pulses ($\Delta t_1$).  
The figure also shows the normalized phase variations in J1 and J2 as a function of the external flux, where each $2\pi$ increment indicates a single SFQ switching event.  
These results demonstrate how the rate of change of the input signal maps directly to the frequency of the SFQ pulses.}
\label{fig:mod1_pulse}
\end{figure}

Figure~\ref{fig:mod1_pulse} illustrates the simulated phase dynamics of Josephson Junctions J1 and J2, normalized to units of $2\pi$.  
As the external magnetic flux $\Phi_{ext}$ varies, the circulating current within the loop changes accordingly.  
When the flux through the loop reaches roughly half a flux quantum ($0.5\Phi_0$), the screening current can no longer compensate for the external field, prompting one of the junctions to switch.  
This transition releases an SFQ pulse and resets the stored magnetic energy.  
A positive slope in the input waveform activates the “Up” junction (J1), while a negative slope causes the “Down” junction (J2) to switch.  
Each $2\pi$ phase slip across a junction corresponds directly to a quantized SFQ emission event.

The SFQ pulse train produced by Modulator-1 confirms that the output frequency is directly proportional to the slope of the input signal.  
When the input waveform rises more steeply, the interval between consecutive pulses decreases, indicating faster pulse generation.  
Conversely, gentler slopes result in fewer pulses that are more widely spaced.  
In the simulated case where the positive input slope $m_1$ is twice the magnitude of the negative slope $m_2$, the resulting time intervals satisfy $\Delta t_2 = 2\Delta t_1$.  
This relationship verifies the inverse proportionality between the rate of change of the input current and the spacing of the SFQ pulses.

\subsubsection{Modulator-2 Simulation Results}

\begin{figure}[!htbp]
\centering
\includegraphics[width=1\linewidth]{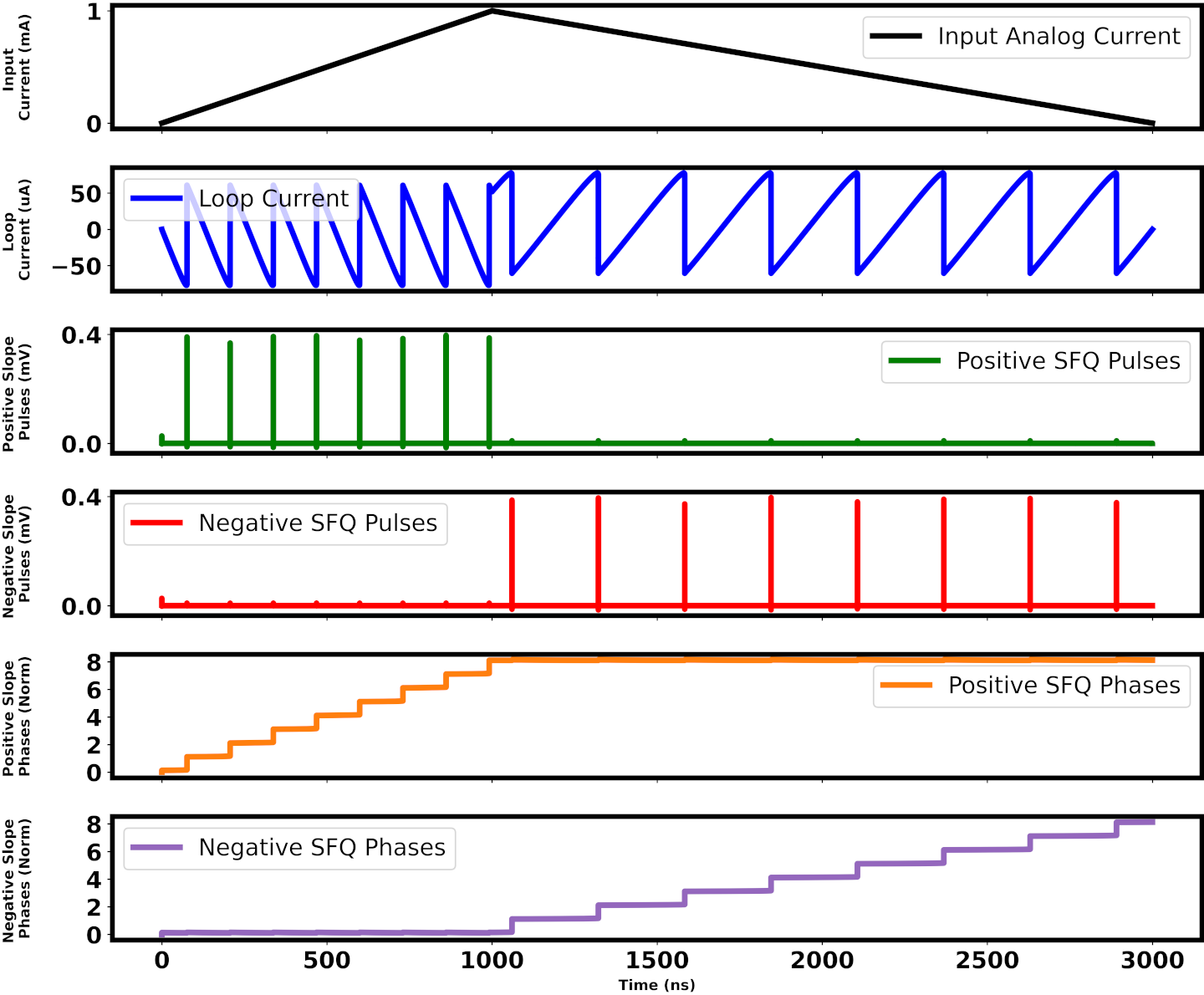}
\caption{Simulation result for Modulator-2.  
The larger loop inductance increases the sensitivity to small magnetic flux variations, resulting in more frequent SFQ switching events.  
In this example, Modulator-2 generates eight SFQ pulses for the same input signal that produces three pulses in Modulator-1.  
The normalized phase traces of J1 and J2 show more frequent $2\pi$ phase transitions, illustrating the enhanced flux resolution of the circuit.  
Although higher sensitivity comes with slightly reduced bandwidth and greater noise susceptibility, Modulator-2 provides finer temporal quantization of the input waveform, demonstrating the tunable balance between sensitivity and speed.}
\label{fig:mod2_pulse}
\end{figure}

Figure~\ref{fig:mod2_pulse} presents the simulated phase evolution of J1 and J2 for Modulator-2, which features the higher loop inductance of 100\,pH.  
Compared to Modulator-1, the same input waveform produces a greater number of phase transitions, showing that the circuit is more responsive to small flux variations.  
This behavior is consistent with the theoretical prediction that a larger $L_{loop}$ enhances sensitivity by lowering the current required to induce one flux quantum.  
In practice, Modulator-2 generates eight SFQ pulses under the same input conditions where Modulator-1 produces only three.  
This confirms that increasing loop inductance results in denser SFQ pulse sequences, providing finer quantization of the input signal.

The corresponding SFQ pulse outputs reveal shorter pulse spacing and higher switching activity for Modulator-2, reflecting its greater flux resolution.  
While this improvement enhances sensitivity, it also introduces trade-offs: the larger loop inductance slightly reduces the modulator’s bandwidth and increases its sensitivity to thermal noise.  
Nonetheless, the simulation clearly demonstrates that Modulator-2 captures smaller signal variations and provides higher temporal precision in the pulse domain.

\subsection{Majority Design}
Due to their high sensitivity, SQUIDs are inherently susceptible to noise disturbances. When noise is superimposed on the input signal, simulations reveal the generation of spurious SFQ pulses that degrade output fidelity and compromise data integrity. The severity of noise-induced distortion depends on the slope of the input signal; signals with smaller slopes exhibit higher susceptibility to noise when tested under varying slope conditions.

\begin{figure}[!htbp]
    \centering
    \includegraphics[width=1\linewidth]{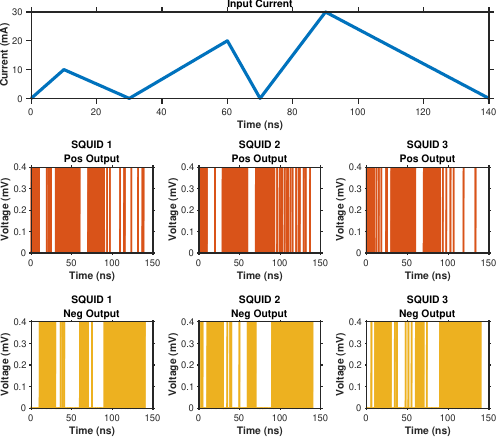}
    \caption{Effect of noise on digital SQUID outputs for different input slopes. Slope values: $m_1 = 1\,\mathrm{mA/ns}$, $m_2 = -0.5\,\mathrm{mA/ns}$, $m_3 = 0.67\,\mathrm{mA/ns}$, $m_4 = -2\,\mathrm{mA/ns}$, $m_5 = 1.5\,\mathrm{mA/ns}$, $m_6 = -0.75\,\mathrm{mA/ns}$. The relation between the absolute values is $|m_2| < |m_6| < |m_1| < |m_3| < |m_5| < |m_4|$. The noisy input is applied to three SQUIDs, and the effect of noise is observed for different slopes. The distortion is more pronounced for lower slope magnitudes.}
    \label{fig:digiSquid3_noise}
\end{figure}

\begin{figure}[!h]
    \centering
    \begin{subfigure}{0.48\linewidth}
        \centering
        \includegraphics[width=\linewidth]{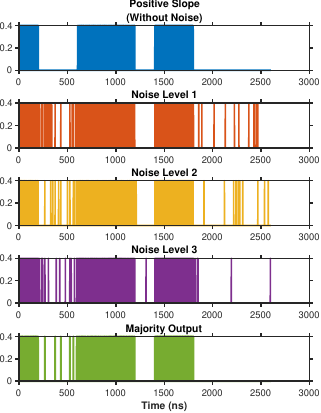}
        \caption{Positive-slope input}
        \label{fig:digiSquid3_majority_a}
    \end{subfigure}
    \hfill
    \begin{subfigure}{0.48\linewidth}
        \centering
        \includegraphics[width=\linewidth]{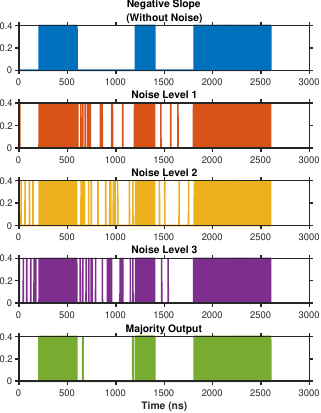}
        \caption{Negative-slope input}
        \label{fig:digiSquid3_majority_b}
    \end{subfigure}
    \caption{Effect of noise on three digital SQUID outputs: (a) positive-slope case; (b) negative-slope case. The figure compares the noise-free output, the outputs of three noisy SQUIDs, and the resulting majority output. The majority result closely replicates the noise-free waveform, demonstrating effective noise suppression.}
    \label{fig:digiSquid3_majority}
\end{figure}

To address noise effects, we explored the triple modular redundancy (TMR) approach, utilizing the simplicity and compact design of a digital SQUID cell, which consists of an inductive loop and two Josephson junctions. Prior to implementing the hardware, we developed a software-based model to simulate three digital SQUIDs receiving the same noisy input. We then post-processed their outputs using a majority-voting algorithm to determine the correct output. The simulation results, depicted in Fig.~\ref{fig:digiSquid3_majority}, demonstrate that majority voting effectively reduces noise-induced errors. The combined output closely aligns with the noise-free reference, validating the effectiveness of the TMR approach as a preprocessing step for noise reduction.

After conducting preliminary evaluations, a hardware-level majority-voting architecture was developed to function within the SFQ domain. The proposed circuit consists of three identical digital SQUIDs operating in parallel, all driven by the same noisy input signal. The outputs from these SQUIDs are directed into an SFQ voting logic block, which generates an output pulse only when at least two SQUIDs produce coincident pulses. If fewer than two SQUIDs generate pulses, the block suppresses any isolated spurious events. While this approach increases the hardware footprint by tripling the number of SQUIDs, it significantly improves noise immunity and overall signal reliability.

\begin{figure}[!htbp]
    \centering
    \includegraphics[width=1\linewidth]{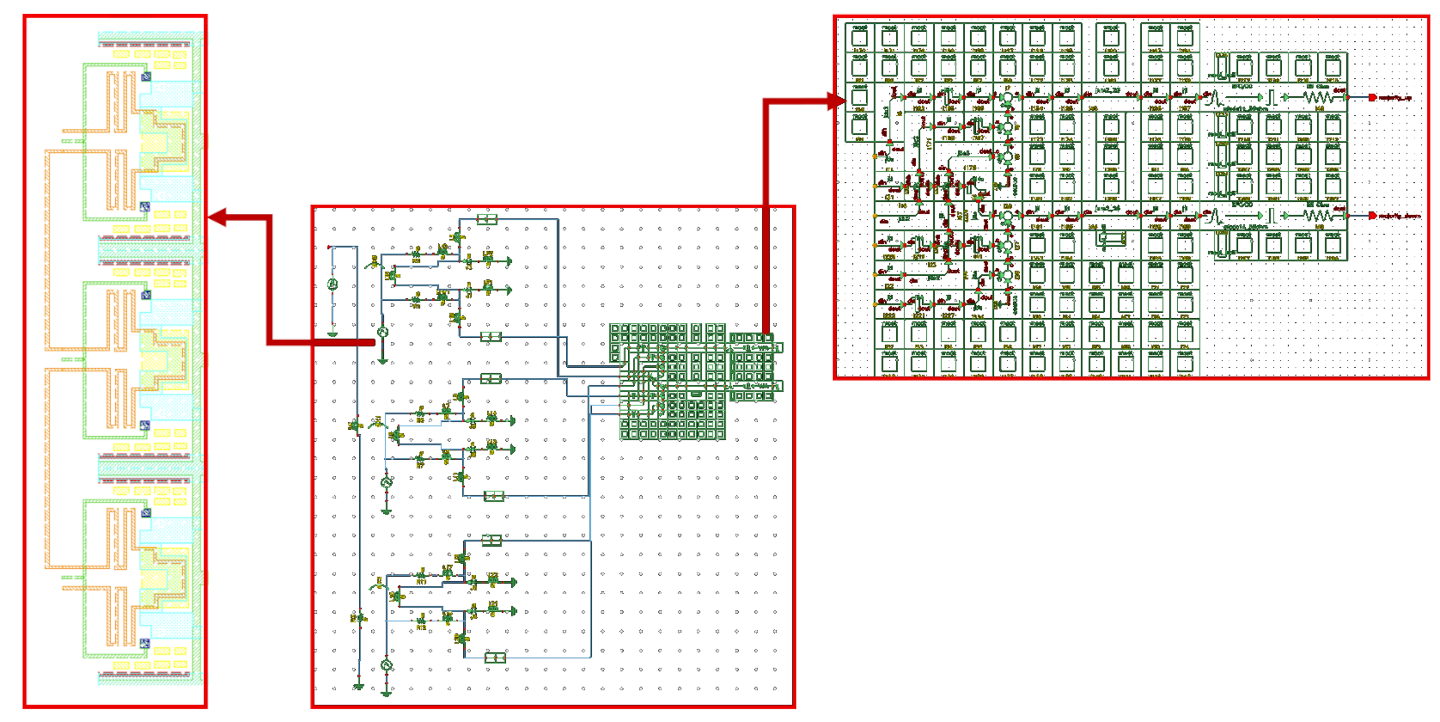}
    \caption{Schematic of the proposed majority-voting circuit. Three digital SQUIDs feed their outputs to a JJ-Soma thresholding unit, which produces an SFQ pulse when two or more coincident pulses are detected.}
    \label{fig:digiSquid_majority_sch}
\end{figure}

The schematic of the proposed circuit is shown in Fig.~\ref{fig:digiSquid_majority_sch}. Due to the asynchronous nature of the digital SQUID outputs, a JJ-Soma circuit—functionally equivalent to an asynchronous AND gate or a thresholding unit—was utilized to detect temporally overlapping SFQ pulses from at least two SQUIDs. This setup allows majority detection to occur entirely within the SFQ domain, eliminating the need for global clock synchronization.

\begin{figure}[!htbp]
    \centering
    \includegraphics[width=1\linewidth]{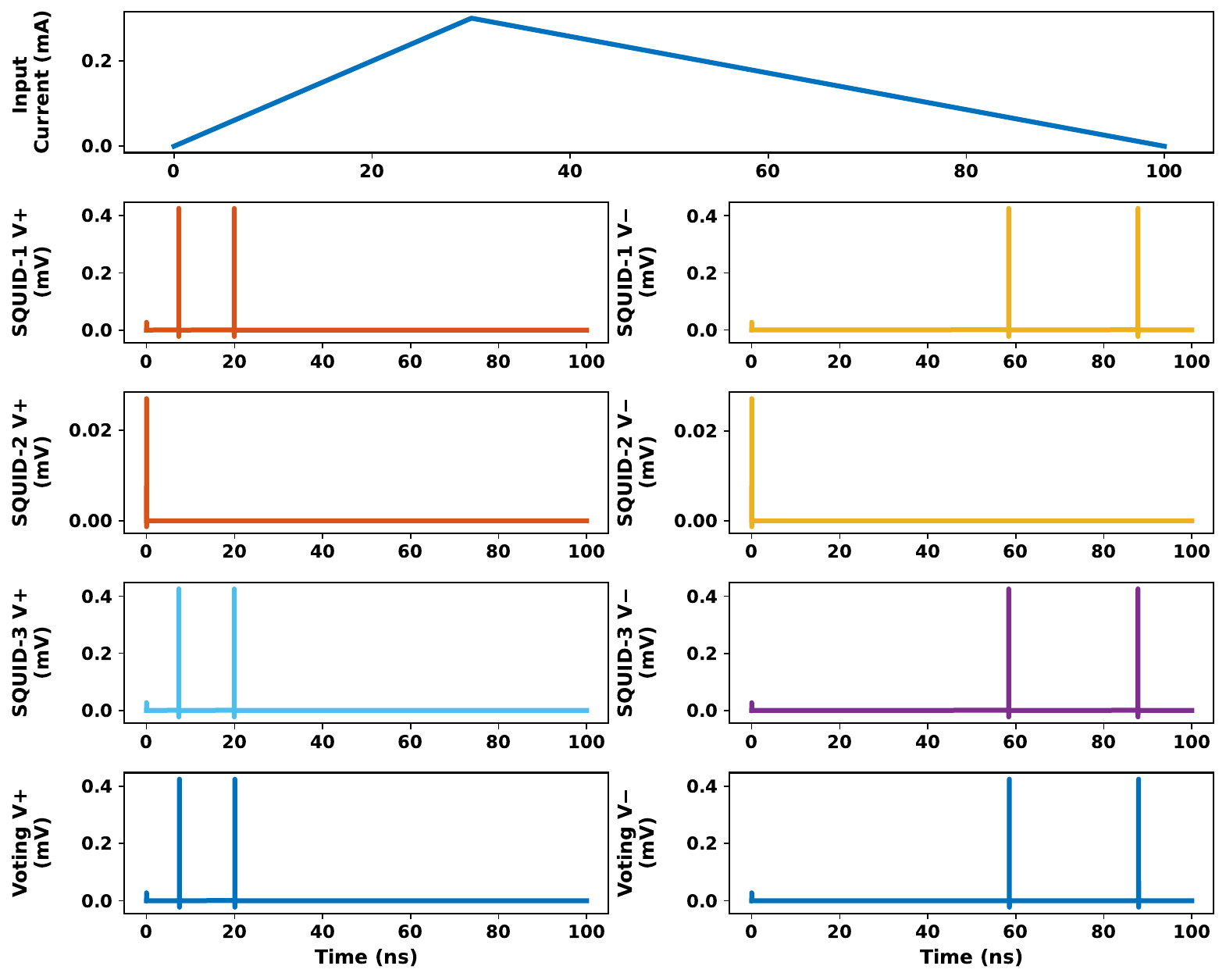}
    \caption{Simulation results of the proposed majority-voting circuit showing improved signal integrity through noise suppression.}
    \label{fig:digiSquid_majority_sim}
\end{figure}

Simulation results presented in Fig.~\ref{fig:digiSquid_majority_sim} confirm the functionality of the proposed majority circuit. The output SFQ pulse is generated when pulses arrive from at least two input channels, demonstrating successful majority detection and effective suppression of noise-induced errors in both positive and negative slope cases.

\subsection{Readout Circuits Design}
The primary advantage of the Digital SQUID is its ability to produce digital outputs in the form of SFQ pulses. To process and interpret these pulses, two distinct digital signal processing circuits were designed: an asynchronous flash counter for high-speed pulse counting and a synchronous cumulative circuit for high-sensitivity, noise-tolerant operation.

\subsubsection{Asynchronous Flash Counter}
The asynchronous flash counter is designed to process the SFQ pulse outputs from the Digital SQUID directly. Given the SQUID’s inherently asynchronous nature, an asynchronous readout architecture provides high-speed operation without the constraints of a global clock. Two separate flash counters are employed to monitor the SQUID’s positive- and negative-slope outputs independently, enabling rapid detection of the magnitude and direction of signal variations.

\begin{figure}[!htbp]
    \centering
    \includegraphics[width=0.8\linewidth]{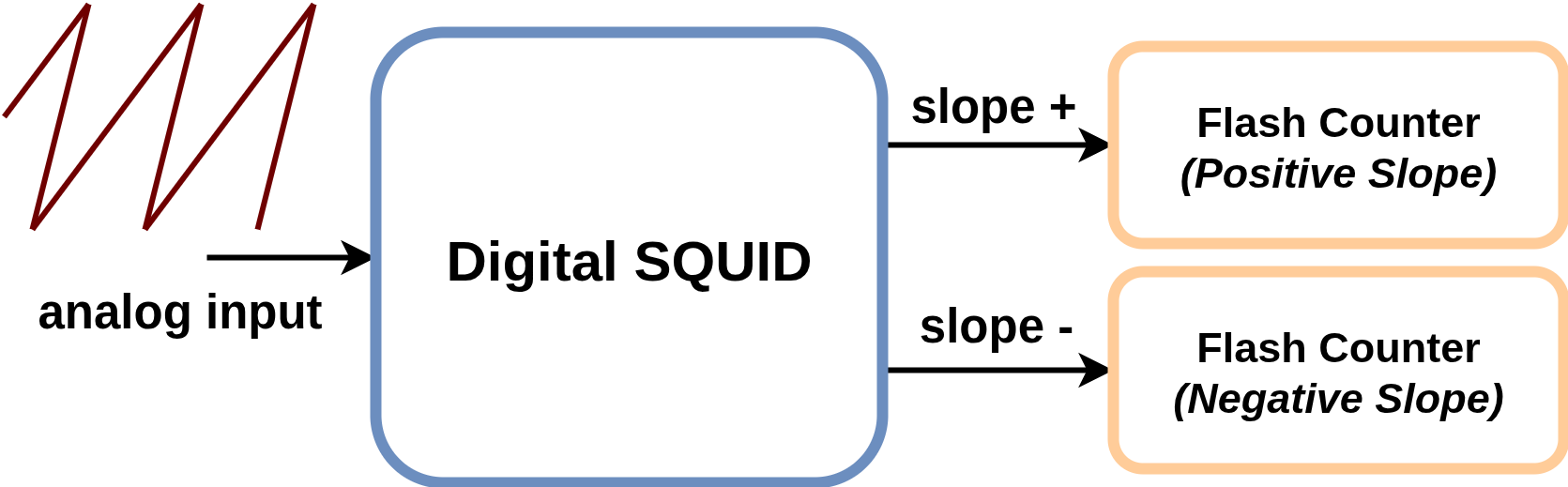}
    \caption{Flash counter block diagram. The positive- and negative-slope outputs of the Digital SQUID are connected to individual asynchronous counters composed of TFF stages.}
    \label{fig:flash_cnt_chip}
\end{figure}

The flash counter exploits the frequency-dividing characteristic of cascaded toggle flip-flops (TFFs). In a 4-bit implementation, four TFFs are connected in series, where the output of each stage serves as both a counter bit and the input for the subsequent stage. The outputs of the Digital SQUID’s positive and negative slopes are fed into the first TFF of each counter chain. To interpret the binary count values, SFQ-to-DC converters are attached to the TFF outputs, providing measurable DC voltages that correspond to the counter's instantaneous state. This asynchronous readout architecture enables compact, high-speed pulse counting, which is compatible with the SQUID’s event-driven behavior.

\subsubsection{Synchronous Cumulative Circuit}
A synchronous cumulative circuit offers an alternative readout strategy that emphasizes sensitivity and noise averaging. Unlike the flash counter, it operates synchronously with a global clock, enabling controlled sampling of SFQ pulses. Although slower, this approach offers improved robustness against transient fluctuations.

\begin{figure}[!htbp]
    \centering
    \includegraphics[width=1\linewidth]{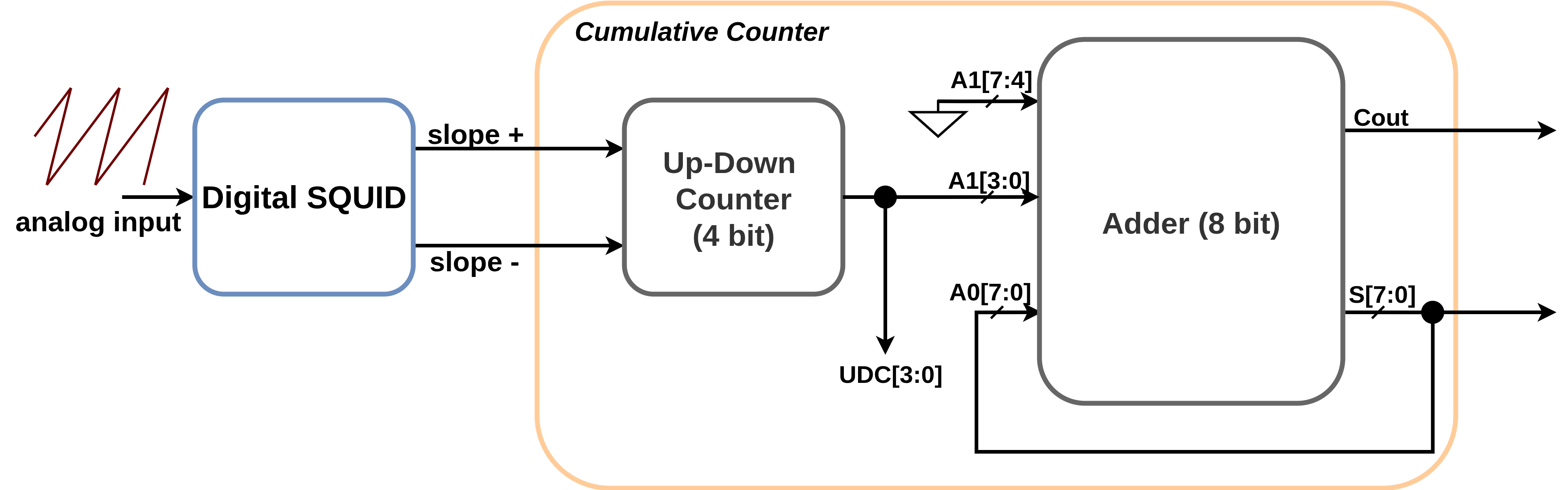}
    \caption{Block diagram of the cumulative digital signal processing circuit. The positive- and negative-slope outputs are applied to an Up–Down Counter, whose output is fed to an adder with a feedback path for cumulative averaging.}
    \label{fig:cumulative_cnt_block}
\end{figure}

As illustrated in Fig.~\ref{fig:cumulative_cnt_block}, the circuit consists of a 4-bit Up–Down Counter that receives the Digital SQUID’s positive- and negative-slope pulses as separate inputs, allowing the system to track both the direction and magnitude of flux variations. The counter output is forwarded to an 8-bit adder, whose output is recursively fed back as one of its inputs in the subsequent clock cycle, forming a cumulative feedback loop. This configuration effectively computes a running sum or average of the recent pulse history, enhancing noise rejection and improving measurement stability. Although the synchronous operation limits speed compared to the flash counter, the feedback-based averaging mechanism makes it particularly suitable for high-sensitivity applications.

The Up–Down Counter is implemented using RSFQ logic cells. The adder is constructed from cascaded T1 gates, each functioning as a 1-bit full adder in RSFQ logic. Eight such T1 stages form an 8-bit adder. The feedback path is implemented using passive transmission line (PTL) cells, which provide ultra-fast feedback propagation with a signal velocity approximately one-third of the speed of light.

\subsection{Readout Circuit Simulation Results}

\subsubsection{Up–Down Counter Simulation}
\begin{figure}[!htbp]
    \centering
    \includegraphics[width=1\linewidth]{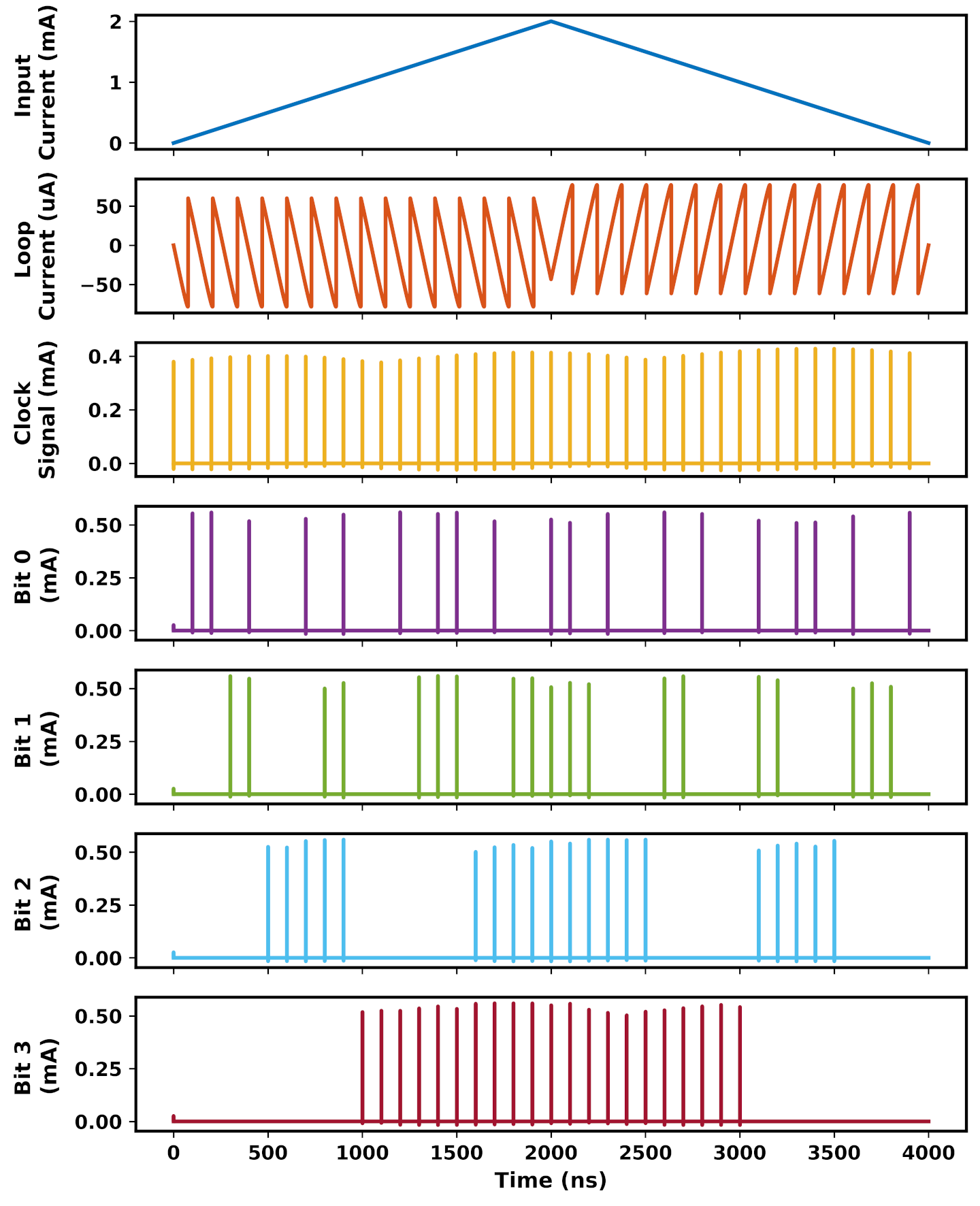}
    \caption{Simulation results of the synchronous Up–Down Counter. The Digital SQUID output pulses drive the counter, which increments or decrements based on pulse polarity, producing 4-bit binary outputs that track the input transitions.}
    \label{fig:updown_sim}
\end{figure}

Simulation results of the Up–Down Counter, driven by the Digital SQUID outputs, are shown in Fig.~\ref{fig:updown_sim}. The figure illustrates the synchronous operation of the counter, where pulses are sampled according to the system clock. Positive input pulses increment the count, while negative pulses decrement it. The resulting four-bit outputs clearly demonstrate the sequential counting behavior—incrementing up to a maximum value and then decreasing as the input polarity reverses.

\subsubsection{Cumulative Counter Simulation}
\begin{figure}[!htbp]
    \centering
    \includegraphics[width=1\linewidth]{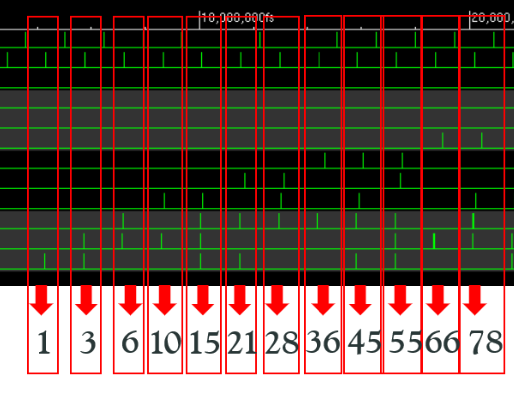}
    \caption{Simulation results of the cumulative readout circuit. Each successive input pulse updates the cumulative sum through feedback, demonstrating the circuit’s ability to compute a running average of the incoming SFQ pulse data.}
    \label{fig:cumulative_sim}
\end{figure}

Figure~\ref{fig:cumulative_sim} presents the simulation results for the cumulative circuit. The circuit behavior confirms continuous accumulation of the Up–Down Counter output with the previous adder state through the feedback loop. Each input pulse modifies the counter output, which is then added to the stored cumulative value, demonstrating the system's averaging and noise-reducing functionality. The cumulative circuit produces a 9-bit output, reflecting extended resolution through accumulation over multiple input cycles.

\section{Experimental Results}
\subsection{Experimental Setup}
\begin{figure}[!htbp]
    \centering
    \begin{subfigure}{0.48\linewidth}
        \centering
        \includegraphics[width=\linewidth]{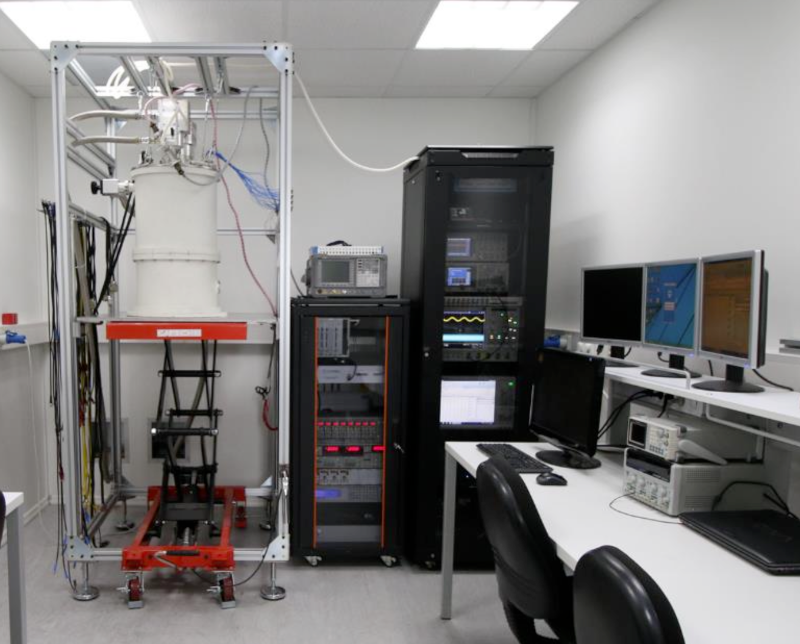}
        \caption{Experimental setup inside the Faraday cage, including the pulse-tube cryocooler, power supplies, amplifiers, and test instrumentation.}
        \label{fig:full_system}
    \end{subfigure}
    \hfill
    \begin{subfigure}{0.48\linewidth}
        \centering
        \includegraphics[width=\linewidth]{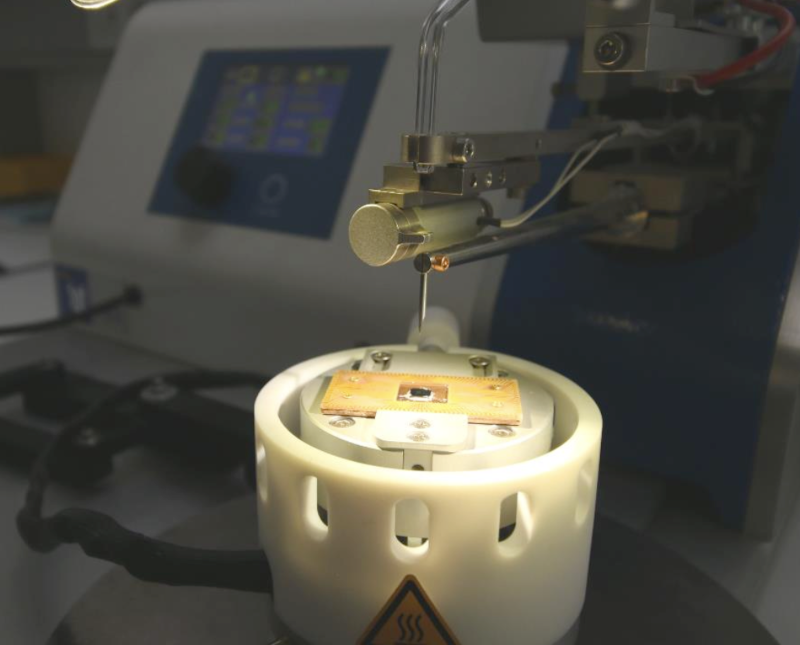}
        \caption{Chip under test during the wire-bonding process using gold wire connections for input/output interfacing.}
        \label{fig:wirebond}
    \end{subfigure}
    \caption{Experimental setup for cryogenic testing:
    (a) Measurement infrastructure inside the Faraday cage; 
    (b) Wire-bonding process for the superconducting chip.}
    \label{fig:full_system_full}
\end{figure}

Testing superconducting circuits requires operation at cryogenic temperatures well below this threshold. Although liquid helium cooling at 4.2~K has traditionally been used, its high cost and limited availability make it less practical for repeated experiments. In this work, a two-stage closed-cycle cryogenic system was employed, capable of achieving a base temperature of 4.2 K after an initial precooling stage at approximately 35 K. The overall experimental setup is illustrated in Fig.~\ref{fig:full_system_full}.

The cryogenic platform employs a pulse-tube cryocooler to minimize mechanical vibrations, which can otherwise disrupt the operation of sensitive superconducting circuits. To further attenuate residual vibrations and improve thermal isolation, an intermediate damping stage is inserted between the cold head and the sample mount. The system is equipped with a LabVIEW-based control interface that enables real-time monitoring and adjustment of temperature, vacuum level, and compressor parameters.

For accurate and low-noise measurements, careful consideration was given to wiring and shielding. Phosphor-bronze cables were used for DC bias and power lines to reduce thermal conduction and maintain stable operation. RF-shielded coaxial lines were employed for signal transmission, particularly for high-frequency inputs. Input bias and clock signals were generated using a digital waveform generator and passed through cryogenic low-pass filters before reaching the device under test. A HYPRES programmable current source supplied precise bias currents for the circuit operation. Electrical interconnections between the chip and the package were made using gold wire bonds to ensure low-resistance, thermally stable contacts.

Separate DC bias lines were provided for the DC/SFQ, SFQ/DC, and main power rails to isolate functional blocks and minimize cross-interference. An external signal generator delivered clock signals through a bias-tee configuration. The SFQ output signals from the chip, located inside the cryostat, were amplified by a low-noise differential amplifier with adjustable gain and subsequently monitored using a high-bandwidth oscilloscope and a digital logic analyzer. This configuration enabled time-domain characterization of SFQ pulse waveforms and validation of correct logic behavior under cryogenic operation.

\subsection{Experimental Results}

The experimental validation began with testing the digital components of the system before full integration with the analog front-end. The initial tests focused on verifying the operation of the Toggle Flip-Flop (TFF) counters, followed by integration with the Digital SQUID modulator to evaluate the complete asynchronous counting behavior.

\begin{figure}[!htbp]
    \centering
    \includegraphics[width=1\linewidth]{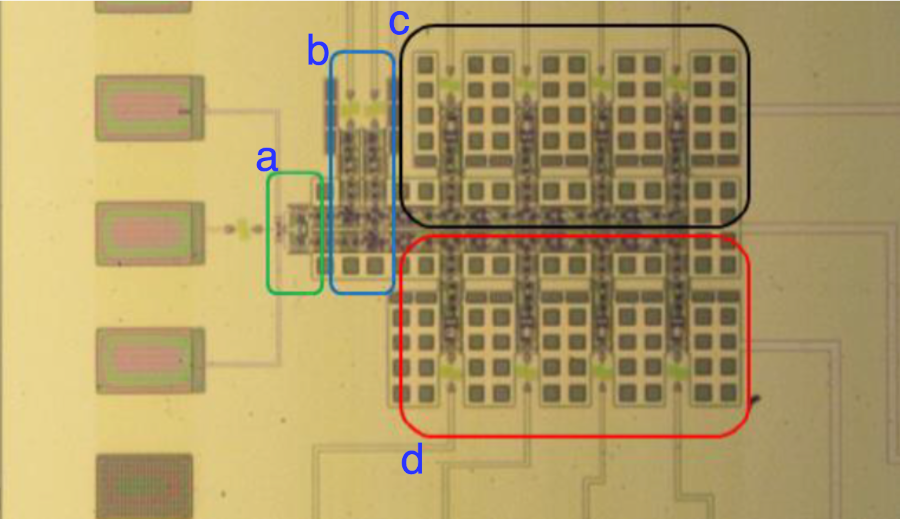}
    \caption{Sections of the chip under test for validating the functionality of the modulator and TFF counters individually. 
    (a) Digital-SQUID modulator; (b) monitoring node to observe the modulator output and independently drive the TFF counters; (c) 4-bit up counter for positive slope pulses; (d) 4-bit down counter for negative slope pulses.}
    \label{fig:tff_test}
\end{figure}

\begin{figure}[!htbp]
    \centering
    \includegraphics[width=1\linewidth]{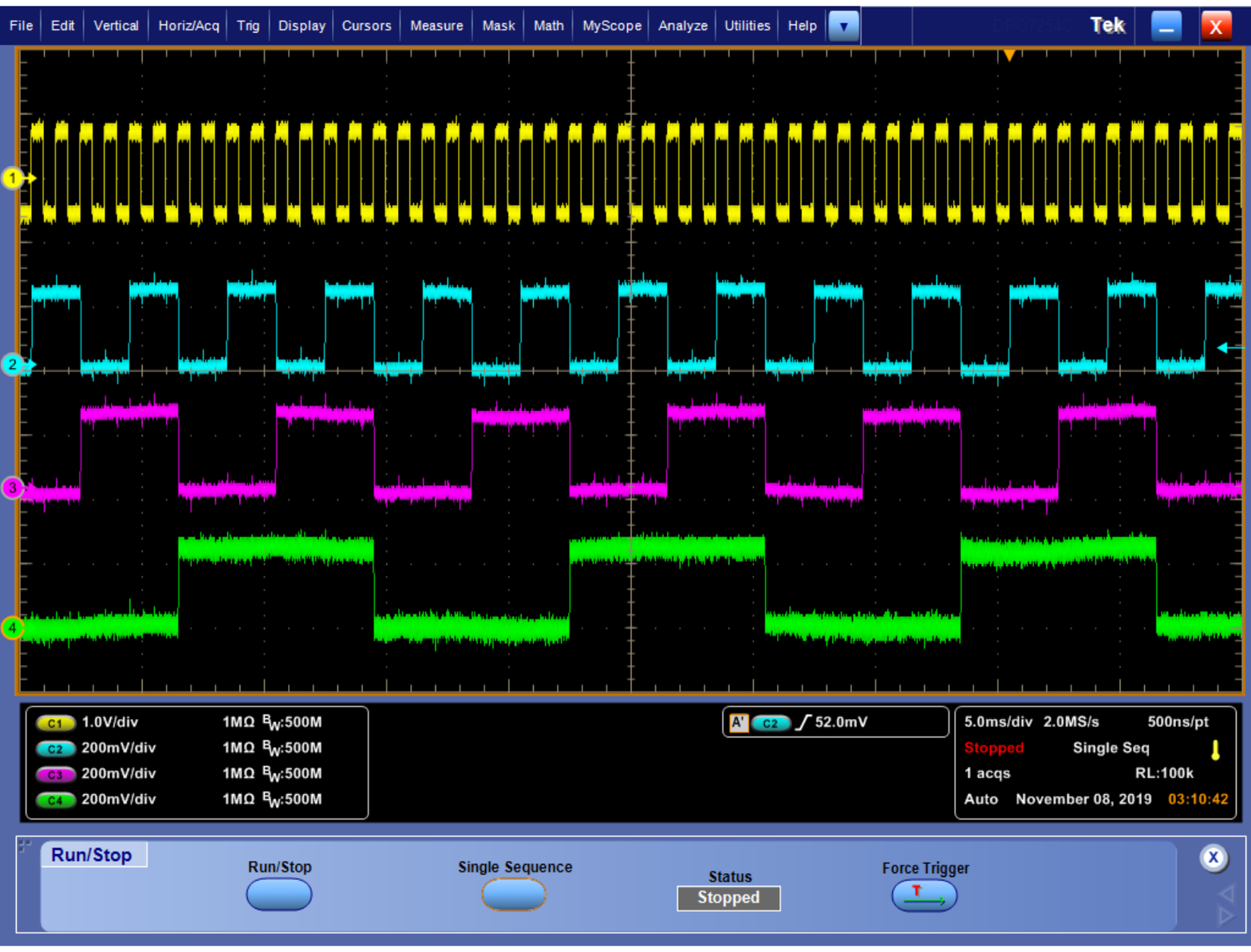}
    \caption{TFF counter test results. The input signal (top trace) and the subsequent three TFF outputs demonstrate the divide-by-two characteristic of cascaded TFFs.}
    \label{fig:tff_test_result}
\end{figure}

The testing sequence began with the standalone digital section. Once the counter functionality was verified, the analog Digital SQUID modulator was integrated. The TFF counter successfully demonstrated its frequency-dividing behavior, as shown in Fig.~\ref{fig:tff_test_result}. The top trace corresponds to the input signal, and the subsequent traces represent the TFF outputs, confirming correct divide-by-two operation at each stage. This verified the correct functionality of the individual TFF cells and their cascading behavior.

\begin{figure}[!htbp]
    \centering
    \includegraphics[width=1\linewidth]{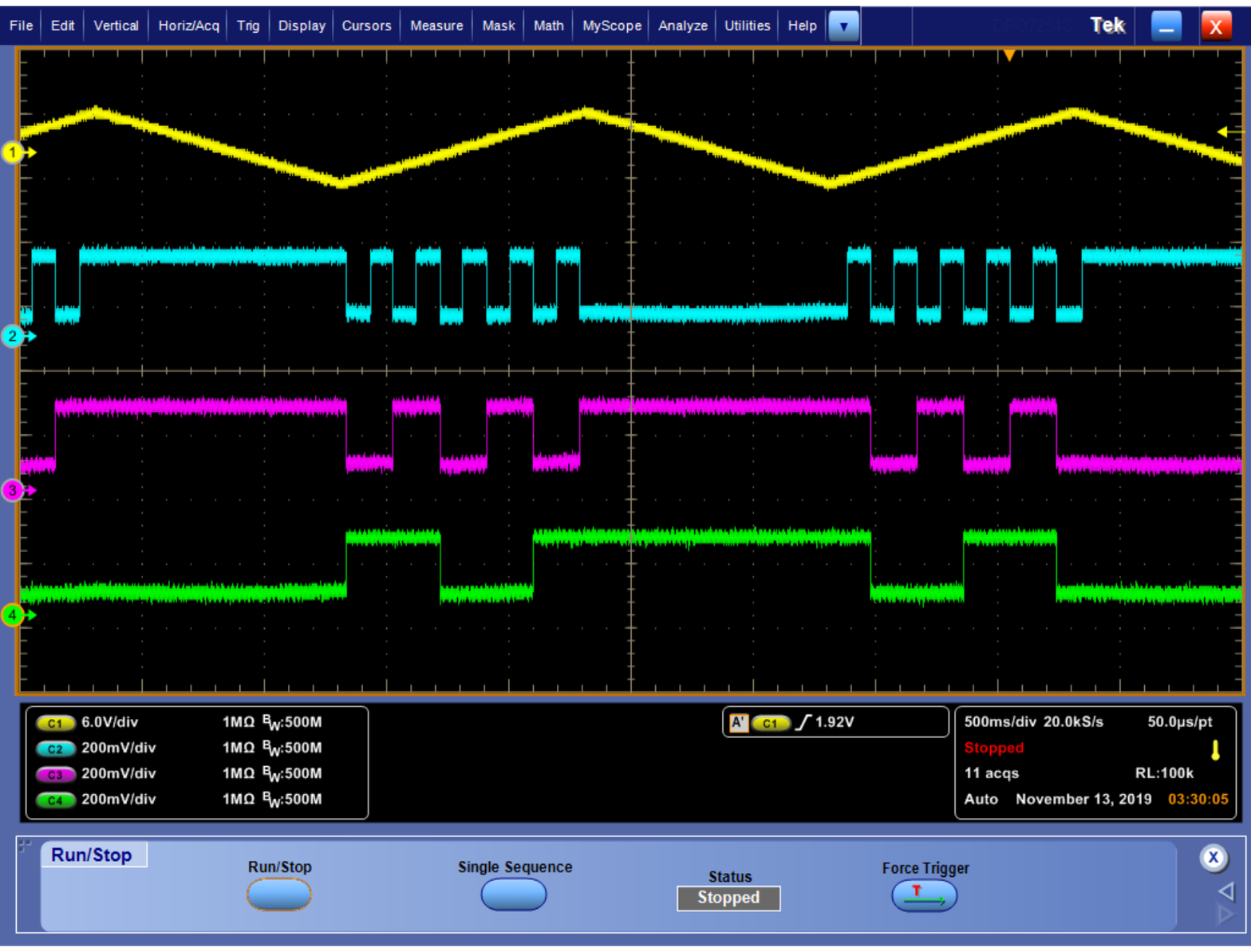}
    \caption{Positive-slope counter measurement results. The positive portion of the input waveform activates the up counter, verifying the correct operation of both the Digital SQUID and the asynchronous flash counter.}
    \label{fig:mod_pos_counter}
\end{figure}

After validating the digital block, the Digital SQUID modulator was activated, and its outputs were fed to the TFF counters. A triangular current waveform with alternating positive and negative slopes was applied to the modulator input. The positive-slope portions of the waveform triggered the up counter, while the negative slopes triggered the down counter. The measured results in Fig.~\ref{fig:mod_pos_counter} confirm correct counting of the SFQ pulses corresponding to positive slope intervals.

\begin{figure}[!htbp]
    \centering
    \includegraphics[width=1\linewidth]{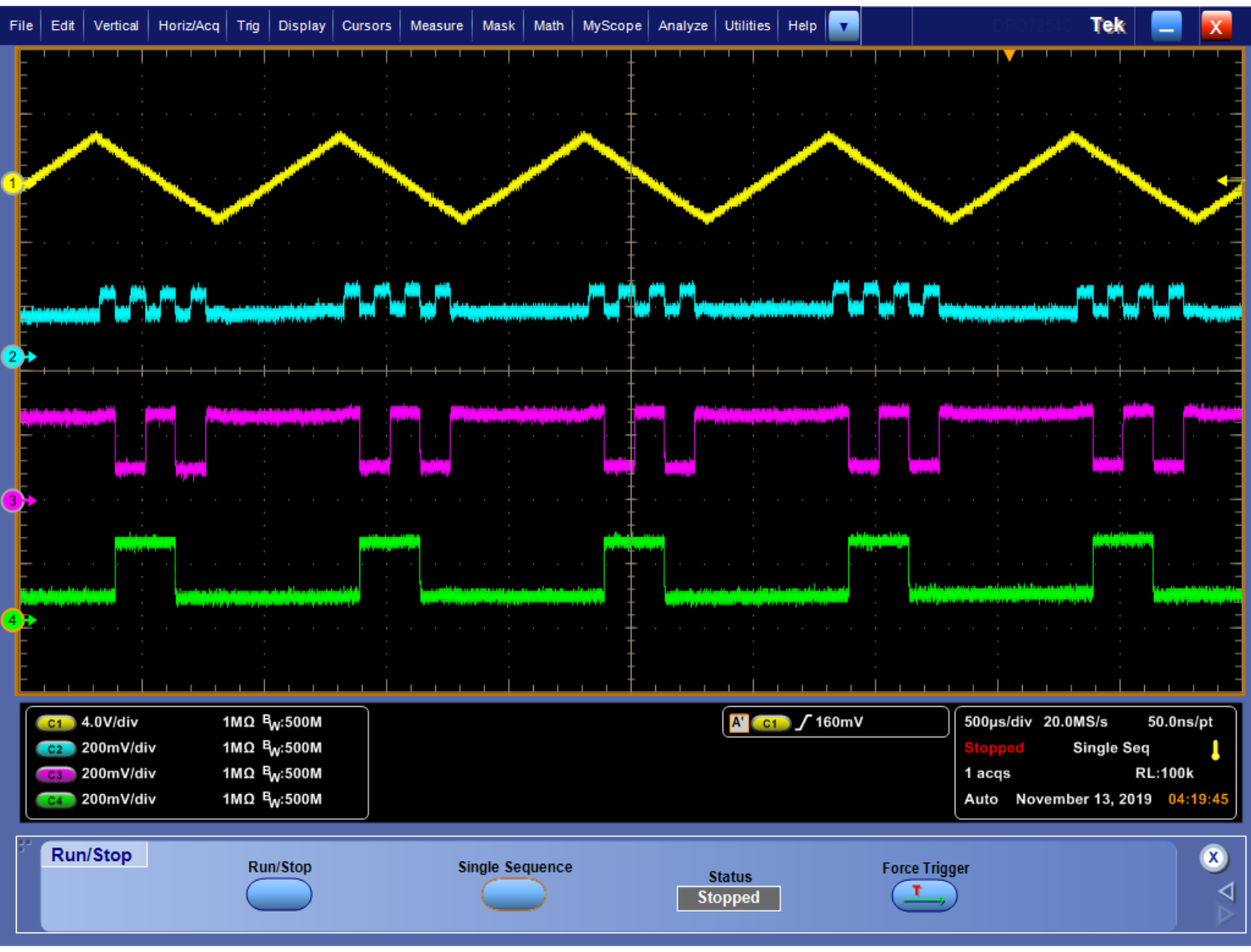}
    \caption{Negative-slope counter measurement results. The down counter correctly responds to the negative-slope segments of the input, validating the bidirectional operation of the Digital SQUID–counter system.}
    \label{fig:mod_neg_counter}
\end{figure}

Similarly, Fig.~\ref{fig:mod_neg_counter} shows the response of the down counter when driven by the negative-slope portions of the same input waveform. The output confirms that the counter increments appropriately in response to negative-slope events, validating the correct directional operation of the Digital SQUID and counter combination.

\begin{figure}[!htbp]
    \centering
    \begin{subfigure}{0.48\linewidth}
        \centering
        \includegraphics[width=\linewidth]{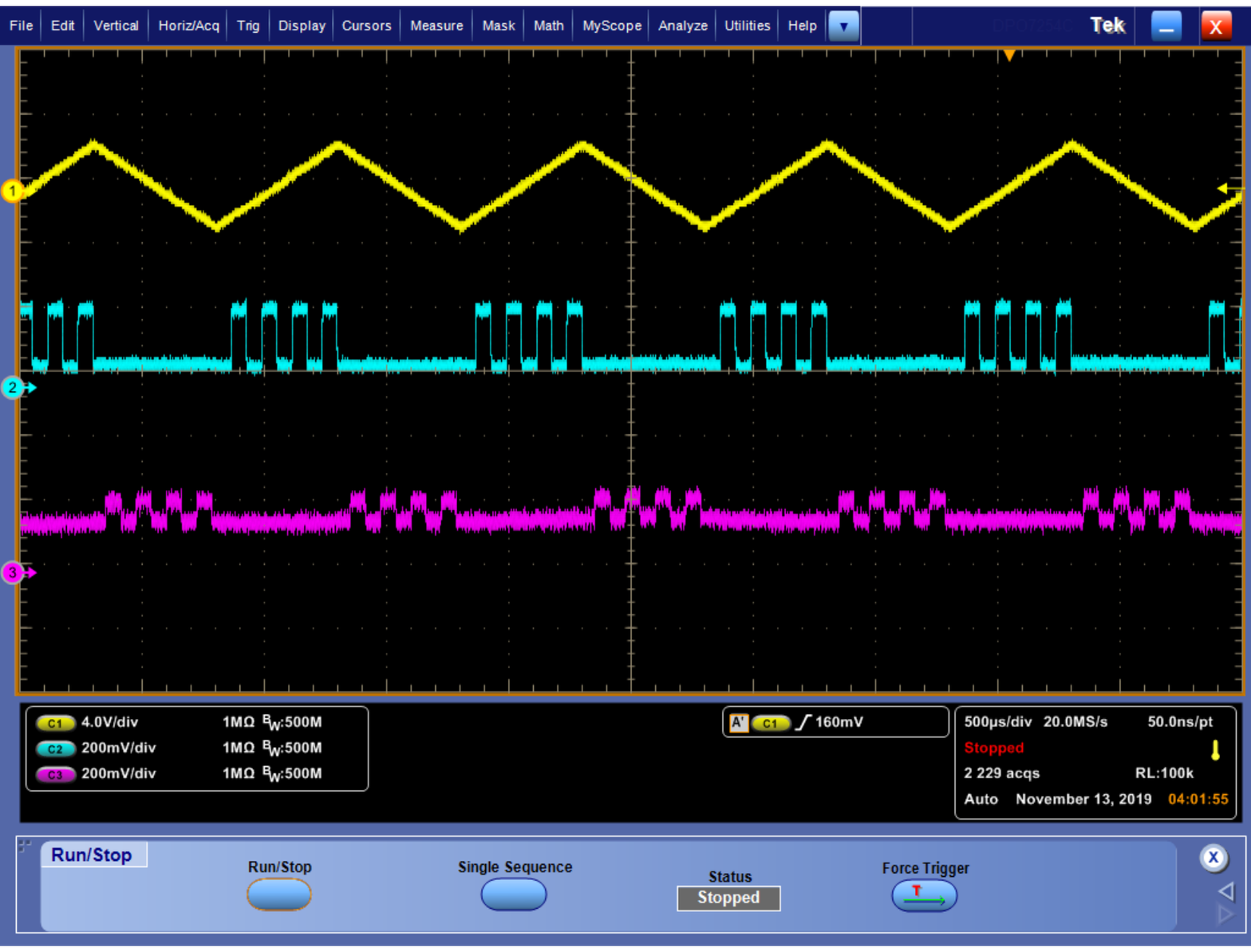}
        \caption{Bit0: positive and negative slope counters}
    \end{subfigure}
    \hfill
    \begin{subfigure}{0.48\linewidth}
        \centering
        \includegraphics[width=\linewidth]{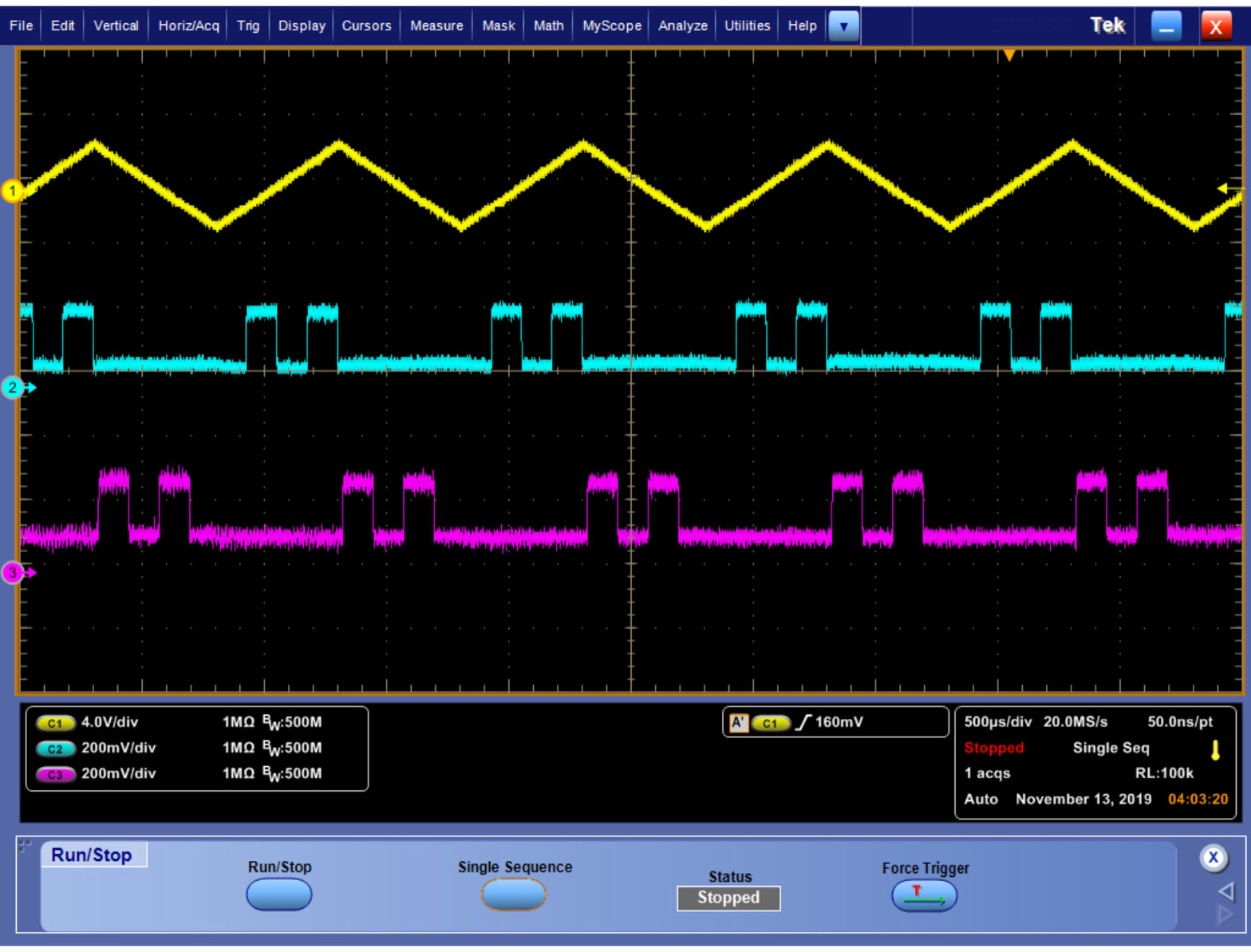}
        \caption{Bit1: positive and negative slope counters}
    \end{subfigure}

    \vspace{2mm}

    \begin{subfigure}{0.48\linewidth}
        \centering
        \includegraphics[width=\linewidth]{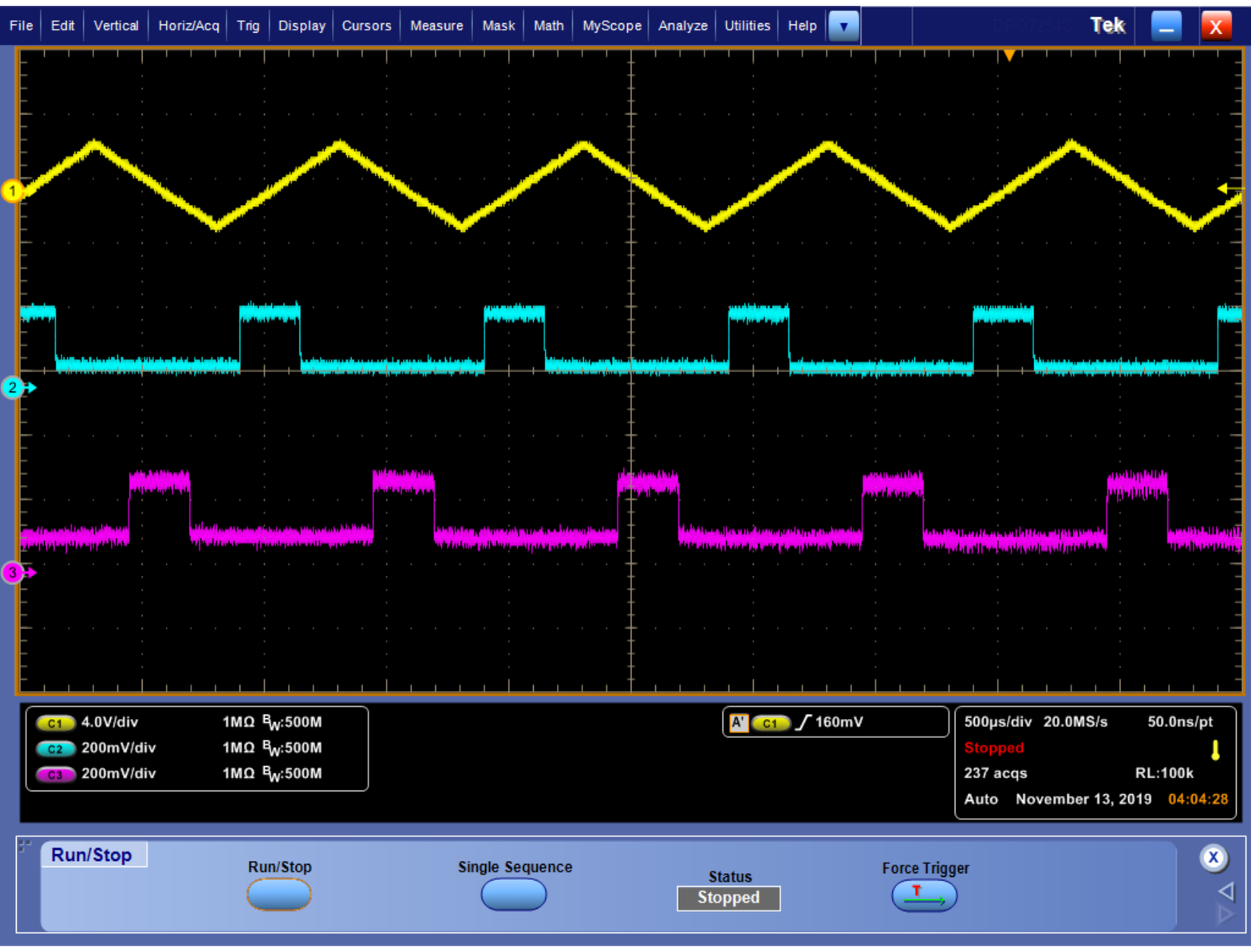}
        \caption{Bit2: positive and negative slope counters}
    \end{subfigure}
    \hfill
    \begin{subfigure}{0.48\linewidth}
        \centering
        \includegraphics[width=\linewidth]{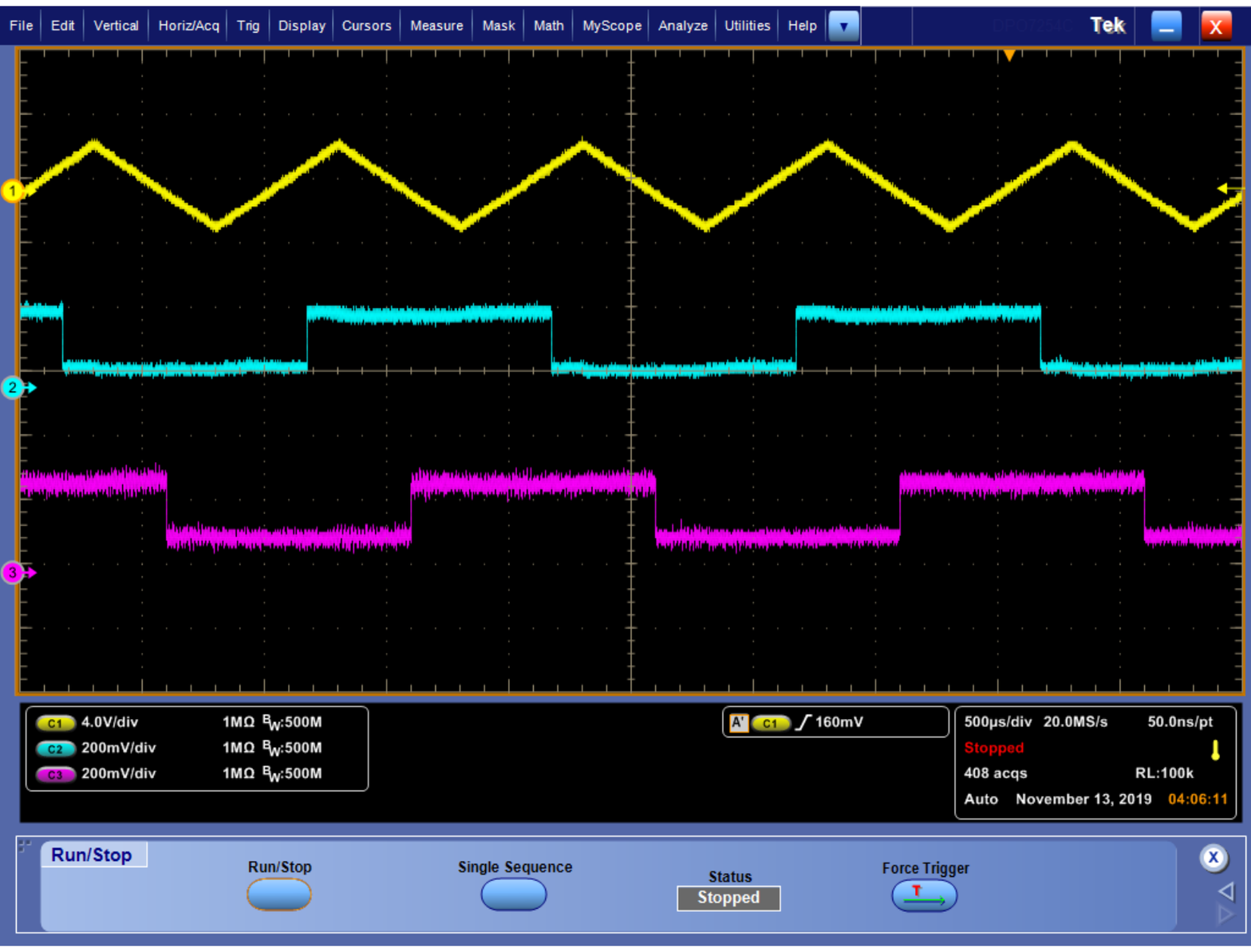}
        \caption{Bit3: positive and negative slope counters}
    \end{subfigure}

    \caption{Counter comparison for identical slope inputs. All bits exhibit the same activity in both positive- and negative-slope counters, confirming consistent operation across both counting paths.}
    \label{fig:2x2grid}
\end{figure}

To further confirm functional symmetry, identical slope values were applied to both the positive- and negative-slope inputs, and the outputs of the corresponding counters were compared. As shown in Fig.~\ref{fig:2x2grid}, all four counter bits (Bit0–Bit3) exhibit identical activity for both input directions, confirming consistent counting performance across all TFF stages and validating the overall modulator–counter integration.

The uniform activity across all counter bits validates the joint functionality of the Digital SQUID modulator and the TFF-based asynchronous counters, confirming the correct integration of the analog and digital domains.

\begin{figure}[!htbp]
    \centering
    \includegraphics[width=1\linewidth]{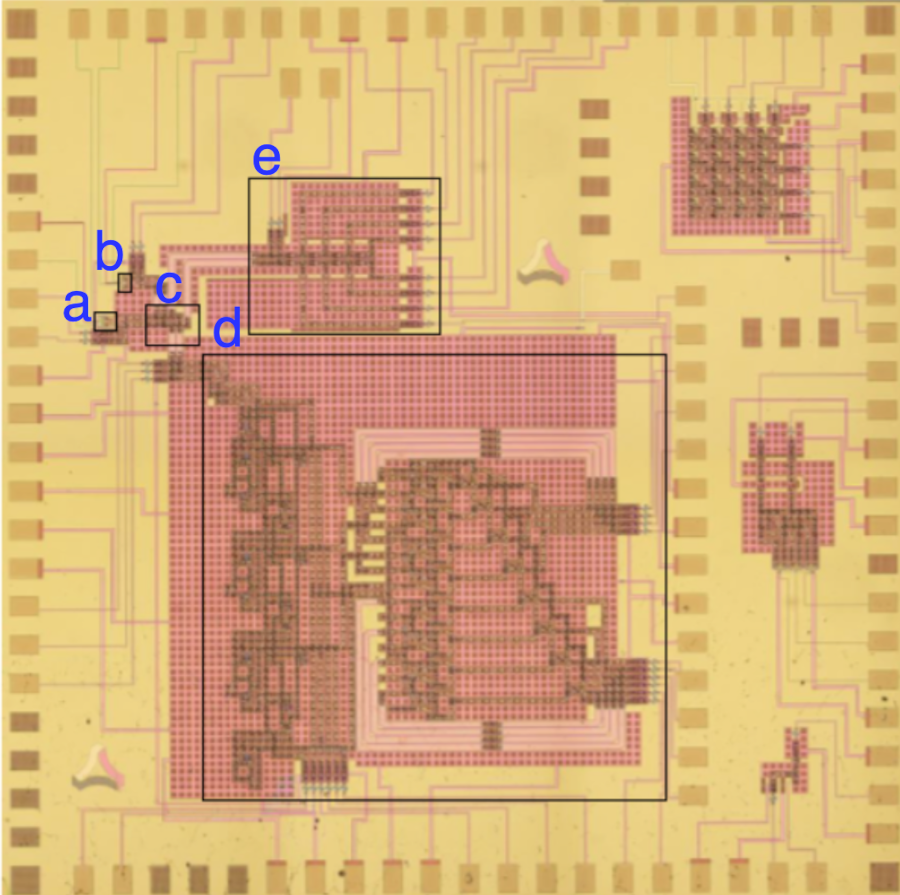}
    \caption{Micrograph of the fabricated chip integrating all circuit components: (a) Digital SQUID~1 (high sensitivity), (b) Digital SQUID~2 (low sensitivity), (c) selection bias lines, (d) asynchronous TFF counters, and (e) synchronous cumulative counter.}
    \label{fig:chip_full}
\end{figure}

Following the successful validation of the subsystem, a fully integrated chip was fabricated, incorporating all design components on a single die. The complete circuit includes two Digital SQUID modulators with different sensitivities, asynchronous up/down counters, and a synchronous cumulative counter for averaging and noise suppression. The layout of the full design is shown in Fig.~\ref{fig:chip_full}.

% \section{Applications}
% Superconducting magnetometers, such as SQUIDs, are utilized as both sensors and amplifiers, with extensive applications including earthquake detection, magnetoencephalography, magnetocardiography, solid-state physics, biomagnetism, and archaeology. With the proposed modulator, we can accurately read the cryogenic sensor signal at a very high rate. With preprocessing done at the cryogenic stage using SFQ circuits, the data integrity of the sensor will remain intact. In addition to SQUIDs, photon detectors such as the superconductor stripline photon detector (SSPD), the kinetic inductance detector (KID), and microwave detector antennas can be connected to the modulator, allowing data to be extracted at high frequencies and with quantum-level accuracy. This can be used in high-fidelity quantum readout and communication.

\section{Conclusion}
This work presents the design and experimental validation of a fully integrated superconductor analog-to-digital converter system based on digital SQUID modulators. Two modulators with different sensitivities were developed to convert analog input currents into SFQ pulse streams, while two DSP circuits—an asynchronous flash counter and a synchronous cumulative counter were designed to interpret these pulses on-chip. The asynchronous counter provides high-speed operation, whereas the synchronous cumulative circuit offers enhanced noise resilience through feedback-based averaging. To mitigate noise-induced errors, particularly at low input slopes, a majority-voting circuit employing three modulators was introduced as a pre-filtering mechanism, effectively improving signal reliability. All circuits, including the modulators and DSP units, were integrated on a single chip, and part of the design was experimentally verified. The successful operation of the TFF counters and their combined functionality with the modulators confirms the feasibility of the proposed architecture. The demonstrated system provides a low-noise and high-speed foundation for superconducting analog-to-digital conversion. Its flexibility to select between high-speed asynchronous and high-sensitivity synchronous processing paths makes it a promising platform for future cryogenic computing and sensing applications.

\ackn This work has been funded by ANR and TUBITAK under the Bosphorus program, with project number 117F266.\\
Grammarly AI was used for proofreading the text.

\section*{References}
\bibliographystyle{iopart-num}
%\bibliography{references}

\begin{thebibliography}{10}
\expandafter\ifx\csname url\endcsname\relax
  \def\url#1{{\tt #1}}\fi
\expandafter\ifx\csname urlprefix\endcsname\relax\def\urlprefix{URL }\fi
\providecommand{\eprint}[2][]{\url{#2}}

\bibitem{5972910}
Gupta D, Inamdar A~A, Kirichenko D~E, Kadin A~M and Mukhanov O~A 2011 Superconductor analog-to-digital converters and their applications {\em 2011 IEEE MTT-S International Microwave Symposium\/} pp 1--4

\bibitem{razmkhahBook}
{Razmkhah, Sasan and Febvre, Pascal} 2023 {Superconducting Quantum Electronics} {\em {Beyond-CMOS}\/} (ISTE \& WILEY) chap~8, pp 295--391 ISBN 9781394228713

\bibitem{holmes2021cryogenic}
Holmes D~S 2021 Cryogenic electronics and quantum information processing {\em 2021 IEEE International Roadmap for Devices and Systems Outbriefs\/} (IEEE) pp 1--93

\bibitem{razmkhah2024challenges}
Razmkhah S, Aviles R~S, Li M, Gupta S, Beerel P~A and Pedram M 2024 Challenges and unexplored frontiers in electronic design automation for superconducting digital logic {\em 2024 Design, Automation \& Test in Europe Conference \& Exhibition (DATE)\/} (IEEE) pp 1--6

\bibitem{4277786}
Vernik I~V, Kirichenko D~E, Filippov T~V, Talalaevskii A, Sahu A, Inamdar A, Kirichenko A~F, Gupta D and Mukhanov O~A 2007 {\em IEEE Transactions on Applied Superconductivity\/} {\bf 17} 442--445

\bibitem{mizugaki20149}
Mizugaki Y, Takahashi Y, Shimada H and Maezawa M 2014 {\em Electronics Letters\/} {\bf 50} 1637--1639

\bibitem{guelaz2009wide}
Guelaz R, Desgreys P and Loumeau P 2009 Wide-band sigma--delta adc design in superconducting technology {\em Languages for Embedded Systems and their Applications: Selected Contributions on Specification, Design, and Verification from FDL’08\/} (Springer) pp 101--112

\bibitem{phillips1987superconducting}
Phillips R~R, Sandell R~D and Silver A~H 1987 Superconducting analog-to-digital converter with bidirectional counter uS Patent 4,646,060

\bibitem{mukhanov20117}
Mukhanov O 2011 {\em 100 Years of Superconductivity\/}  440

\bibitem{9153926}
Tanaka M, Kozaka M, Kamiya K, Üşenmez K, Aydoğan E~C, Razmkhah S, Bozbey A and Fujimaki A 2021 {\em IEEE Transactions on Applied Superconductivity\/} {\bf 31} 1--6

\bibitem{9373955}
Bozbey A, Aydoğan E~C, Üşenmez K, Razmkhah S, Tanaka M and Fujimaki A 2021 {\em IEEE Transactions on Applied Superconductivity\/} {\bf 31} 1--5

\bibitem{Fujimaki1988_25}
Fujimaki N, Tamura H, Imamura T and Hasuo S 1988 {\em IEEE Transactions on Electron Devices\/} {\bf 35} 2412--2418

\bibitem{Radparvar1994_26}
Radparvar M 1994 {\em IEEE Transactions on Applied Superconductivity\/} {\bf 4} 87--91

\bibitem{Radparvar1997_27}
Radparvar M and Rylov S~V 1997 {\em IEEE Transactions on Applied Superconductivity\/} {\bf 7} 3682--3685

\bibitem{Uhlmann1999_28}
Uhlmann F~H, Lange S, Khabipov M and Meyer H~G 1999 {\em Physica C: Superconductivity\/} {\bf 326--327} 72--78

\bibitem{Rylov1991_29}
Rylov S~V 1991 {\em IEEE Transactions on Magnetics\/} {\bf 27} 2431--2434

\bibitem{Yuh1995_30}
Yuh P~F and Rylov S~V 1995 {\em IEEE Transactions on Applied Superconductivity\/} {\bf 5} 2129--2132

\bibitem{Semenov2003_31}
Semenov V~K 2003 {\em IEEE Transactions on Applied Superconductivity\/} {\bf 13} 747--750

\bibitem{Gupta2001_32}
Gupta D and Radparvar M 2001 {\em IEEE Transactions on Applied Superconductivity\/} {\bf 11} 1261--1264

\bibitem{Reich2005_33}
Reich T, Ortlepp T and Uhlmann F~H 2005 {\em IEEE Transactions on Applied Superconductivity\/} {\bf 15} 304--307

\bibitem{Myoren2011_34}
Myoren H, Kimimoto Y, Terui K and Taino T 2011 {\em IEEE Transactions on Applied Superconductivity\/} {\bf 21} 387--390

\bibitem{Tsuga2013_35}
Tsuga Y, Yamanashi Y and Yoshikawa N 2013 {\em IEEE Transactions on Applied Superconductivity\/} {\bf 23} 1601405

\bibitem{Myoren2019_36}
Myoren H, Okabe K, Matsunawa R, Naruse M and Taino T 2019 {\em IEEE Transactions on Applied Superconductivity\/} {\bf 29} 1--4

\bibitem{Reich2007_37}
Reich T, Ortlepp T and Uhlmann F~H 2007 {\em IEEE Transactions on Applied Superconductivity\/} {\bf 17} 746--749

\bibitem{hidaka2021fabrication}
Hidaka M and Nagasawa S 2021 {\em IEICE Transactions on Electronics\/} {\bf 104} 405--410

\bibitem{5740577}
Sarreshtedari F, Hosseini M, Razmkhah S, Mehrany K, Kokabi H, Schubert J, Banzet M, Krause H~J and Fardmanesh M 2011 {\em IEEE Transactions on Applied Superconductivity\/} {\bf 21} 3442--3446

\bibitem{Razmkhah_2021}
Razmkhah S, Bozbey A and Febvre P 2021 {\em Superconductor Science and Technology\/} {\bf 34} 045013 \urlprefix\url{https://dx.doi.org/10.1088/1361-6668/abdedb}

\bibitem{Shnyrkov_2023}
Shnyrkov V~I, Shapovalov A~P, Lyakhno V~Y, Dumik A~O, Kalenyuk A~A and Febvre P 2023 {\em Superconductor Science and Technology\/} {\bf 36} 035005 \urlprefix\url{https://dx.doi.org/10.1088/1361-6668/acb10e}

\bibitem{zimmerman1966coherent}
Zimmerman J, Cowen J and Silver A 1966 {\em Applied Physics Letters\/} {\bf 9} 353--355

\bibitem{clarke2006squid}
Clarke J and Braginski A~I 2006 {\em The SQUID handbook: Applications of SQUIDs and SQUID systems\/} (John Wiley \& Sons)

\bibitem{5634076}
Haverkamp I, Mielke O, Kunert J, Stolz R, Meyer H~G, Toepfer H and Ortlepp T 2011 {\em IEEE Transactions on Applied Superconductivity\/} {\bf 21} 705--708

\bibitem{yilmaz2019flux}
Yilmaz U, Razmkhah S, Bozbey A and Febvre P 2019 Flux-coupled asynchronous hybrid squid operating in a closed-cycle gm cooler {\em 14th European Conference on Applied Superconductivity (EUCAS 2019)\/}

\end{thebibliography}

\providecommand{\newblock}{}

\end{document}